\documentstyle[12pt,axodraw]{article}

\newcommand{\mysection}{\setcounter{equation}{0}\section}
\newcommand{\Slash} {\slash \!\!\!}

\begin{document}
\vskip 0.2cm
\hfill{YITP-SB-03-58}
\vskip 0.2cm
\vskip 0.2cm
\centerline{\large\bf {NNLO corrections to massive lepton-pair 
production }}
\centerline{\large\bf {in longitudinally polarized proton-proton collisions}}
\vskip 0.4cm
\centerline {\sc V. Ravindran}
\centerline{\it Harish-Chandra Research Institute,}
\centerline{\it Chhatnag Road, Jhusi,}
\centerline{\it Allahabad, 211019, India.}
\vskip 0.2cm
\centerline {\sc J. Smith 
\footnote{partially supported
by the National Science Foundation grant PHY-0098527.}
}
\centerline{\it C.N. Yang Institute for Theoretical Physics,}
\centerline{\it State University of New York at Stony Brook,
New York 11794-3840, USA.}
\vskip 0.2cm
\centerline {\sc W.L. van Neerven }
\centerline{\it Instituut-Lorentz}
\centerline{\it University of Leiden,}
\centerline{\it PO Box 9506, 2300 RA Leiden,}
\centerline{\it The Netherlands.}
\vskip 0.2cm
\centerline{October 2003}
\vskip 0.2cm
\centerline{\bf Abstract}
\vskip 0.3cm
We present the full next-to-next-to-leading order (NNLO) coefficient
functions for the polarized cross section $d\Delta \sigma/dQ$ for the Drell-Yan 
process $p + p\rightarrow l^+l^- + 'X'$. 
Here $'X'$ denotes any inclusive hadronic 
state and $Q$ represents the invariant mass of the lepton pair.
All QCD partonic subprocesses have been included provided the lepton pair 
is created by a virtual photon, which is a valid approximation 
for $Q<50~{\rm GeV}$.
Unlike the differential distribution w.r.t. transverse momentum
the dominant subprocess for the integrated cross section is given by 
$q+\bar q \rightarrow \gamma^* + 'X'$ and its higher order corrections so 
that massive lepton pair production provides us with
an excellent tool to measure the polarized anti-quark densities.
Our calculations are carried out using the method of 
$n$-dimensional regularization by making a special choice for the
$\gamma_5$ matrix. We give predictions for double
longitudinal spin asymmetry measurements at the RHIC.

\vfill

\mysection{Introduction}
With the advent of the RHIC at BNL we have a new facility to study the 
spin structure of the proton, (for a review on the potential of the RHIC 
see \cite{busa}), which supplements the existing polarized lepton-hadron
machines.  Polarized proton-proton collisions with a very high luminosity
and a maximum centre of mass energy of $\sqrt s=500~{\rm GeV}$ 
will provide us with many more details about spin distributions
than possible with the existing lepton-hadron machines, which give very 
little information about the polarized gluon and sea-quark parton densities.
If we limit ourselves to the Drell-Yan process where the initial state
hadrons are longitudinally polarized then the gluon initiated subprocess 
dominates the quark process in $d^3\Delta \sigma/dQ^2/dy/dp_T$
at a transverse momentum $p_T>Q/2$ and fixed rapidity $y$
\cite{bgk1}, \cite{bgk2}. This result, however, 
depends on the parton densities involved. For a study of the
cross section $d^3\Delta \sigma/dQ^2/dy/dp_T$ and its dependence on
the parton density set we refer to the complete NLO calculation in \cite{rsn1}
(the non-singlet part was already done in \cite{chco1} and \cite{chco2}).
However for the totally integrated 
cross section $d\Delta \sigma/dQ$ and for $p + p \rightarrow l^+l^- + `X'$ 
the sea 
quark initiated process dominates the reaction for the whole phase space.
Therefore this process provides us with an excellent tool to measure the
sea-quark densities. This process was calculated up to next-to-leading 
order (NLO) in \cite{rat} and subsequently confirmed
in \cite{weber}, \cite{mara} \cite{gehr} and \cite{kamal}. It is our goal to add all 
subprocesses which appear in the next-to-next-to-leading order 
(NNLO) contribution to $d\Delta \sigma/dQ$
and to see whether the sea quark dominance persists. 
In order to calculate such a process one has first to specify the scheme
for the $\gamma_5$ matrix if one wants to use $n$-dimensional regularization
and massless quarks. In this paper we use the HVBM prescription of 't Hooft and
Veltman \cite{hove} which is more elaborated upon by Breitenlohner and Maison 
in \cite{brma}. However this prescription like any others requires the
introduction 
of evanescent counter terms before the reaction can be renormalized by the
standard procedure. An example for the HVBM prescription is that the
non-singlet axial vector operator gets renormalized in spite of the fact
that it is conserved. This effect has to be undone by introducing an additional
renormalization constant. This has also to be done for higher spin operators
otherwise their anomalous dimensions do not equal the anomalous 
dimensions of the spin averaged operators. Also the Adler-Bardeen theorem
\cite{adba}
is violated which states that the Adler-Bell-Jackiw anomaly \cite{abj} does not
receive higher order QCD corrections. A scheme which is equivalent to
HVBM is the one given by Akyeampong and Delbourgo \cite{akde}. Here 
the $\gamma_5$
matrix is replaced by the Levi-Civita tensor multiplied by four $\gamma$ 
matrices. If the Feynman and phase space integrals are calculated before
the traces are taken the result is the same as in the complicated
HVBM technique. More details are given in the next section.

Our paper is organized as follows. In Section 2 we introduce our notations
and discuss the technicalities which are involved when using 
Akyeampong-Delbourgo prescription for the $\gamma_5$-matrix. 
In Section 3 we present the NNLO corrections to the coefficient functions
of the polarized DY cross section $d\Delta \sigma/dQ$.
In Section 4 we study the NNLO corrections to polarized DY production 
in proton-proton collisions at the RHIC. Notice that this study is not complete
since the three-loop contributions to the splitting functions are not
available yet so that the corresponding polarized parton densities are not
known.
The long formulae for the NNLO coefficient functions
can be found in Appendix A.


\mysection{Kinematics of the polarized Drell-Yan process and $\gamma_5$
scheme}
The Drell-Yan process proceeds through the following reaction
\begin{eqnarray}
\label{eq2.1}
&& H_1(P_1,S_1)+H_2(P_2,S_2)\rightarrow \gamma^*(q) + 'X'\,,
\nonumber\\
&& \hspace*{52mm}\mid
\nonumber\\
&& \hspace*{54mm}\rightarrow l^+(l_1)+l^-(l_2)
\nonumber\\[2ex]
&& S=(P_1+P_2)^2 \,, \qquad Q^2\equiv q^2 =(l_1+l_2)^2 \,,
\end{eqnarray}
where $H_i$ ($i=1,2$) represent the incoming polarized hadrons carrying
the momenta $P_i$ and spins $S_i$. Further $'X'$ denotes any inclusive
hadronic state which is unpolarized. The lepton pair is represented by
$l^+l^-$ with momenta $l_1$, $l_2$. In this paper we will only consider 
lepton pairs which have a sufficiently small invariant mass $Q$ so that 
the photon dominates in the above reaction and $Z$-boson exchange effects 
can be neglected. 
The cross section is given by
\begin{eqnarray}
\label{eq2.2}
\frac{d\,\Delta \,\sigma^{\rm H_1H_2}}{d\,Q^2}=\frac{4\pi\alpha^2}
{3\,N\,Q^2\,S}\,
\Delta W^{\rm H_1H_2}(\tau,Q^2)\,, \qquad \tau=\frac{Q^2}{S}\,.
\end{eqnarray}
In the QCD improved parton model the hadronic DY structure function
$\Delta W^{\rm H_1H_2}$ is related to the coefficient functions
$\Delta_{ij}$ as follows
\begin{eqnarray}
\label{eq2.3}
\Delta W^{\rm H_1H_2}(\tau,Q^2) &=& \sum_{i,j=q,\bar q,g}
\int_{\tau}^1\, \frac{dx_1}{x_1} \int_{\tau/x_1}^1\,\frac{dx_2}{x_2}\,
\Delta f_i^{\rm H_1}(x_1,\mu^2) \,\Delta f_j^{\rm H_2}(x_2,\mu^2)
\nonumber\\[2ex]
&&\times \Delta_{ij} \left (\frac{\tau}{x_1x_2},Q^2,\mu^2\right )\,.
\end{eqnarray}
In the formula above $\Delta f_i(x,\mu^2)$ ($i=q,\bar q,g$) are 
the polarized parton densities where $\mu$ denotes the 
factorization/renormalization scale and $x$ is the fraction of the 
hadron momentum carried by the parton. The DY partonic structure 
function $\Delta \hat W_{ij}$ is computed from the partonic subprocess
\begin{eqnarray}
\label{eq2.4}
i(p_1,s_1)+j(p_2,s_2)\rightarrow \gamma^* (q) + i_1(k_1) \cdots
i_n(k_n)\,, 
\end{eqnarray}
and it reads
\begin{eqnarray}
\label{eq2.5}
&& \Delta \hat W_{ij}=K_{ij}\,\int d^4q\,\delta(q^2-Q^2)
\nonumber\\[2ex]
&&\times\prod_{i=1}^n
\int \frac{d^3k_i}{(2\pi)^3\,2E_i}\,\delta^{(4)}\left (p_1+p_2
- q-\sum_{j=1}^n k_j\right )
\nonumber\\[2ex]
&&\times |\Delta \hat M_{i+j\rightarrow \gamma^* +i_1 \cdots i_n}|^2 \,,
\end{eqnarray}
where $K_{ij}$ denotes the colour and spin average factors and the 
polarized matrix elements are denoted by $\Delta \hat M$ (when we refer to 
unpolarized
structure functions, matrix elements and parton densities
we drop the $\Delta$). Finally note
that the relation between the parton densities above 
and the parton momentum densities appearing
in the parton density sets in the literature or PDF libraries, which are
denoted by $\Delta f_a^{\rm PDF}(x,\mu^2)$, is given by 
$\Delta f_a^{\rm PDF}(x,\mu^2)=x~\Delta f_a(x,\mu^2)$.

When $|\Delta \hat M_{ij}|^2$ in Eq. (\ref{eq2.5}) is calculated up
to order $\alpha_s^2$ one encounters four partonic subprocesses
which are characterised by the two partons in their initial state.  
In the case of quarks with a mass $m\not =0$ they are given by
\begin{eqnarray}
\label{eq2.6}
&& q(p_1,s_1)+\bar q(p_2,s_2) \rightarrow \gamma^* + 'X'\,,
\nonumber\\[2ex]
&& |\Delta \hat M_{q\bar q}|^2=\frac{1}{4}\,{\bf Tr}\,\left (
\gamma_5{\Slash}s_2\,
({\Slash}p_2-m) \,\tilde M \,\gamma_5{\Slash}s_1\,({\Slash}p_1+m)
\,\tilde M^{\dagger}\right )\,,
\\[2ex]
\label{eq2.7}
&& q_1(\bar q_1)(p_1,s_1)+ g(p_2) \rightarrow \gamma^* + 'X'\,,
\nonumber\\[2ex]
&&|\Delta \hat M_{qg}|^2=\frac{1}{4}\,\epsilon_{\mu\nu\lambda\sigma}\,
\frac{p_2^{\lambda} \,l_2^{\sigma}}{p_2\cdot l_2}\,{\bf Tr}\,\left 
(\tilde M^{\mu}
\,\gamma_5{\Slash}s_1\,({\Slash}p_1\pm m)\, \tilde M^{\nu\dagger}\right )\,,
\\[2ex]
\label{eq2.8}
&& q_1(\bar q_1)(p_1,s_1)+ q_2(\bar q_2)(p_2,s_2) \rightarrow \gamma^* + 'X'
\nonumber\\[2ex]
&&|\Delta \hat M_{q_1q_2}|^2=\frac{1}{4}\,{\bf Tr}\,\left (\gamma_5{\Slash}s_2\,
({\Slash}p_2\pm m) \,\tilde M \,\gamma_5{\Slash}s_1\,({\Slash}p_1\pm m)\,
\tilde M^{\dagger}
\right )\,,
\\[2ex]
\label{eq2.9}
&& g(p_1)+ g(p_2) \rightarrow \gamma^* + 'X'\,,
\nonumber\\[2ex]
&&|\Delta \hat M_{gg}|^2=
\nonumber\\[2ex]
&&\frac{1}{4}\,\epsilon_{\mu_2\nu_2\lambda_2\sigma_2}\,
\frac{p_2^{\lambda_2} \,l_2^{\sigma_2}}{p_2\cdot l_2}\,
\epsilon_{\mu_1\nu_1\lambda_1\sigma_1}\,\frac{p_1^{\lambda_1}
\,l_1^{\sigma_1}}{p_1\cdot l_1}\,{\bf Tr}\,\left (\tilde M^{\mu_1\mu_2}\,
\tilde M^{\nu_1\nu_2\dagger}\right )\,,
\end{eqnarray}
where $\tilde M$ denotes the matrix element which is given by the standard
Feynman rules. Further the symbol ${\bf Tr}$ can represent multiple
traces when the matrix elements are calculated in higher order and for
the reaction in Eq. (\ref{eq2.8}) one must distinguish between $q_1=q_2$ and
$q_1\not =q_2$. 
The spin vectors $s_i$ and the gauge vectors $l_i$ ($i=1,2$) satisfy the 
properties
\begin{eqnarray}
\label{eq2.10}
s_i\cdot p_i=0\,, \qquad s_i\cdot s_i=-1\,, \qquad l_i\cdot l_i=0\,.
\end{eqnarray}
When the (anti-)quark is massless then one has to make the replacements
\begin{eqnarray}
\label{eq2.11}
\gamma_5{\Slash}s_i\,({\Slash}p_i\pm m) \rightarrow \pm\,\gamma_5\,h_i
\,{\Slash}p_i\,,
\end{eqnarray}
where $h_i$ ($i=1,2$) represent the helicities of the incoming (anti-)quarks
and the $+$ and $-$ signs on the right-hand side hold for the quarks and 
anti-quarks respectively.
The definitions above are chosen in such a way that the partonic polarized
structure function satisfies the property
\begin{eqnarray}
\label{eq2.12}
\Delta W_{ij}=W_{ij}(+,+)-W_{ij}(+,-)\,,
\end{eqnarray}
with $+,-$ denoting the helicities of the incoming partons.

The computation of the matrix elements in Eqs. (\ref{eq2.6})-(\ref{eq2.9})
and their virtual corrections reveals divergences which occur when the
momenta over which one integrates tend to zero (infrared), infinity 
(ultraviolet) or collinear to another momentum (collinear). The most popular
way to regularize these singularities is to choose the method of 
$n$-dimensional regularization \cite{hove} in which the space is extended to 
$n$ dimensions. The singularities are represented by pole terms of the
type $(1/\varepsilon)^k$ with $n=4+\varepsilon$. This method is very useful
because it preserves the Ward identities in the case of gauge theories.
However this is no longer the case when the $\gamma_5$ matrix 
and the Levi-Civita tensor appear like in Eqs. (\ref{eq2.6})-(\ref{eq2.9})
or in weak interactions. There is no consistent way to generalize
these two quantities in $n$ dimensions contrary to the ordinary 
matrix $\gamma_{\mu}$ or the metric tensor $g_{\mu\nu}$. In the
literature one has proposed various methods to extend the $\gamma_5$ matrix
and the Levi-Civita tensor to $n$ dimensions but one always needs 
so-called evanescent counter terms \cite{bogi} to restore the Ward identities.
A very popular prescription is the HVBM-scheme which was proposed in
\cite{hove} and generalized in \cite{brma}. 
In this approach the $n$-dimensional gamma-matrices and momenta have to be 
split 
into $4$ and $n-4$ dimensional parts. Therefore also the integrals
over the final state momenta have to split up in the same way. Many NLO
calculations have been done in this scheme (see e.g.
\cite{chco1}-\cite{kamal}). However this approach requires a special
procedure to deal with the gamma-matrix algebra which is not implemented
in the program FORM \cite{form}. Since this program is used in our
calculations we prefer
another prescription for the $\gamma_5$-matrix which is given in \cite{akde}.
It gives the same results as the HVBM-scheme but it is much simpler to 
use in algebraic manipulation programs. Moreover one does not have to split 
up the integrals over the final state momenta and one can simply use the phase
space integrals computed for unpolarized reactions. 
The procedure in \cite{akde} is given by
\begin{itemize}
\item[1.]
Replace the $\gamma_5$-matrix by
\begin{eqnarray}
\gamma_{\mu}\,\gamma_5=\frac{i}{6}\,\epsilon_{\mu\rho\sigma\tau}\,
\gamma^{\rho}\, \gamma^{\sigma}\,\gamma^{\tau}\quad \mbox {or} \quad
\gamma_5=\frac{i}{24}\,\epsilon_{\rho\sigma\tau\kappa}\,\gamma^{\rho}\,
\gamma^{\sigma}\,\gamma^{\tau}\,\gamma^{\kappa} \,.
\nonumber
\end{eqnarray}
\item[2.]
Compute all matrix elements in $n$ dimensions.
\item[3.]
Evaluate all Feynman integrals and phase space integrals in $n$-dimensions.
\item[4.]
Contract the Levi-Civita tensors in four dimensions after the Feynman 
integrals and phase space integrals are carried out.
\end{itemize}
\begin{eqnarray}
\label{eq2.13}
\end{eqnarray}
Note that the contraction in four dimensions only applies to Lorentz
indices which are present in the Levi-Civita tensors. The last step in
(\ref{eq2.13}) requires that one first has to apply tensorial reduction to all
integrals. For simple expressions this takes the form 
\begin{eqnarray}
\label{eq2.14}
\int \frac{d^n~k}{(2\pi)^n}\,k^{\mu}\,k^{\nu}\,f(k,p_1,p_2)&=&A_{00}(n)\,
g^{\mu\nu} +A_{11}(n)\,p_1^{\mu}\,p_1^{\nu}+A_{22}(n)\,p_2^{\mu}\,p_2^{\nu}
\nonumber\\[2ex]
&& +A_{12}(n)\,p_1^{\mu}\,p_2^{\nu}+A_{21}(n)\,p_2^{\mu}\,p_1^{\nu}\,,
\nonumber\\[2ex]
\int dPS^{(2)} \,k^{\mu}\,k^{\nu}\,f(k,p_1,p_2)&=&B_{00}(n)\,g^{\mu\nu}
+B_{11}(n)\,p_1^{\mu}\,p_1^{\nu}+B_{22}(n)\,p_2^{\mu}\,p_2^{\nu}
\nonumber\\[2ex]
&&+B_{12}(n)\,p_1^{\mu}\,p_2^{\nu}+B_{21}(n)\,p_2^{\mu}\,p_1^{\nu}\,,
\nonumber\\[2ex]
\end{eqnarray}
where the coefficients $A_{ij},B_{ij}$ depend on $n=4 + \varepsilon$.
However there exists an alternative method. If the matrix element only contains
one $\gamma_5$ like in $qg$ or even none as in $gg$ one can apply a different
rule. One replaces $\gamma_5$ according to 1 in Eq. (\ref{eq2.13}) 
and one contracts the Levi-Civita tensors in $n$ dimensions before
the Feynman integrals and phase space integrals are carried out.
Further one factor of $1/2$ in Eqs. (\ref{eq2.7}) and (\ref{eq2.9})
is replaced by $1/(n-2)(n-3)$. The result is the same as given by
the method in Eq. (\ref{eq2.13}).

The HVBM-scheme or the prescription above automatically
reproduces the Adler-Bell-Jackiw anomaly (ABJ) \cite{abj} but it
violates the Ward identity for the non-singlet axial vector current
and the Adler-Bardeen theorem \cite{adba}. In order to obtain the correct 
renormalized quantities one therefore needs to invoke additional counter terms 
\cite{larin}, which are called evanescent \cite{bogi} since they do not occur 
for $n=4$.
This procedure was used to obtain the NLO anomalous dimensions
for the spin operators which determine the evolution of the parton spin
densities see \cite{mene}, \cite{vogel}.  
Evanescent counter terms are also needed in cases where collinear divergences
show up like in partonic cross sections (see e.g. \cite{neerv}). This is 
characteristic for the HVBM scheme as well as any other approach.
The reason is that in
$n$-dimensional regularization there exists a one-to-one correspondence
between the ultraviolet divergences occurring in partonic operator matrix 
elements and the collinear divergences appearing in partonic cross sections 
provided both quantities are of twist two type \cite{neerv}. In \cite{masm}
we have calculated all evanescent counter terms for the HVBM-scheme
up to second order. For later convenience we will express them in the
unrenormalized coupling constant
\begin{eqnarray}
\label{eq2.15}
Z_{qq}^{5,{\rm NS},+}&=&\delta(1-x)+\hat a_s\,S_\varepsilon\,
\left (\frac{Q^2}{\mu^2}\right )^{\varepsilon/2}\,\Bigg [z^{(1)}_{qq}
+\varepsilon \bar z^{(1)}_{qq}\Bigg ]+\hat a_s^2\,S_\varepsilon^2\,\left (\frac
{Q^2}{\mu^2}\right )^\varepsilon\,\Bigg[
\nonumber\\[2ex]
&&-\frac{1}{\varepsilon}\,\beta_0\,z^{(1)}_{qq}+ z^{(2),{\rm NS},+}_{qq}
-2\,\beta_0\,\bar z^{(1)}_{qq}\Bigg ]\,,
\nonumber\\[2ex]
a_s&=&\frac{\alpha_s(\mu^2)}{4\pi}\,,
\end{eqnarray}
\begin{eqnarray}
\label{eq2.16}
Z_{qq}^{5,{\rm NS},-}=-\hat a_s^2\,S_\varepsilon^2\,\left (\frac{Q^2}{\mu^2}
\right )^\varepsilon \,\Bigg [z^{(2),{\rm NS},-}_{qq} \Bigg ]\,,
\end{eqnarray}
\begin{eqnarray}
\label{eq2.17}
Z_{qq}^{5,{\rm PS}}=\hat a_s^2\,S_\varepsilon^2\,\left (\frac{Q^2}{\mu^2}
\right )^\varepsilon \,\Bigg [z^{(2),{\rm PS}}_{qq}\Bigg ] \,,
\end{eqnarray}
where $\beta_0$ represents the lowest order coefficient in the beta-function.
Note that we have to take $\ln Q^2/\mu^2$ terms into account otherwise
it is impossible to relate $\Delta \hat W^{\rm NS}_{q\bar q}$ to the unpolarized
expression $\hat W^{\rm NS}_{q\bar q}$. This follows from the definition
of $Z_{qq}^{5,{\rm NS},+}$
\begin{eqnarray}
\label{2.18}
\left (Z_{qq}^{5,{\rm NS},+}\right )^2=-\frac{\Delta \hat W_{q\bar q}^{\rm NS}
(\hat a_s,Q^2/\mu^2)}{\hat W_{q\bar q}^{\rm NS}(\hat a_s,Q^2/\mu^2)}
\end{eqnarray}
Explicitly we can write
\begin{eqnarray}
\label{eq2.19}
Z_{qq}^{5,{\rm NS},+}(x) &=& \delta(1-x) +  \hat a_s\,S_{\varepsilon}\,\left (
\frac{Q^2}{\mu^2}\right )^{\varepsilon/2}\, C_F\, \Bigg [ 
- 8 (1-x) + \varepsilon \Big \{ 
\nonumber\\[2ex]
&& - 8(1-x)\ln(1-x) + 4 (1 - x)\ln x + 4 - 4 x  \Big \} \Bigg ]
\nonumber\\[2ex]
&& +  \hat a_s^2\, S_{\varepsilon}^2\,\left (\frac{Q^2}{\mu^2}
\right )^\varepsilon\,  \Bigg [ C_F^2 \Big \{ -16(1-x) - (16 + 8 x)\ln x
\nonumber\\[2ex]
&& + 16(1-x)\ln x \ln(1-x)  \Big \}
\nonumber\\[2ex]
&& + C_A C_F \Big \{ \frac{1}{\varepsilon} \Big ( \frac{88}{3}(1-x) \Big )
-\frac{856}{9}(1 - x) + 8(1-x)\zeta(2)
\nonumber\\[2ex]
&& +\frac{176}{3}\,(1-x)\,\ln (1-x)+(- \frac{168}{3}+32\,x)\ln x -4(1-x)\ln^2 x \Big \}
\nonumber\\[2ex]
&& + n_f C_F T_f \Big \{
- \frac{1}{\varepsilon}\Big ( \frac{32}{3}(1-x) \Big ) + \frac{176}{9} (1 - x)
+16\,(1-x)\ln x 
\nonumber\\[2ex]
&&-\frac{64}{3}\,(1-x)\,\ln (1-x)\Big \}\Bigg ]\,,
\end{eqnarray}
\begin{eqnarray}
\label{eq2.20}
Z_{qq}^{5,{\rm NS},-}&=&-\hat a_s^2\,S_\varepsilon^2\,\left (\frac{Q^2}{\mu^2}
\right )^\varepsilon\,\Big (C_F^2-\frac{1}{2}C_A C_F
\Big )\, \Bigg [   8(1+x) \Big ( 4 {\rm Li}_2(-x) 
\nonumber\\[2ex]
&&+  4\ln x \ln (1+x) + 2\zeta(2) - \ln^2 x - 3\ln x \Big ) - 56(1-x) \Bigg ]\,,
\nonumber\\[2ex]
\end{eqnarray}
\begin{eqnarray}
\label{2.21}
Z_{qq}^{5,{\rm PS}}&=&\hat a_s^2\,S_\varepsilon^2\,\left (\frac{Q^2}{\mu^2}
\right )^\varepsilon\,n_f\,C_F\,T_f\,\Bigg [16(1-x)+
8(3-x)\ln x + 4(2+x) \ln^2 x\Bigg ]\,.
\nonumber\\[2ex]
\end{eqnarray}
A comparison with previous calculations \cite{neerv}, \cite{masm} shows
that the evanescent counter terms are universal except for the $Q^2/\mu^2$ terms
and the $\varepsilon$ part of the order $\hat a_s$ contribution.

\mysection{Corrections up to NNLO to polarized Drell-Yan production}
In this section we treat the various subprocesses which contribute to
the Drell-Yan reaction up to NNLO.
The Born reaction is given by the process
\begin{eqnarray}
\label{eq3.1}
q + \bar q \rightarrow \gamma^*\,.
\end{eqnarray}
The first order corrections are given by the following subprocesses. The 
bremsstrahlung reaction to the Born process
\begin{eqnarray}
\label{eq3.2}
q + \bar q \rightarrow \gamma^* + g\,,
\end{eqnarray}
plus the one-loop correction to the Born process. Further we have a new process
with a gluon in the initial state 
\begin{eqnarray}
\label{eq3.3}
g + q(\bar q) \rightarrow \gamma^* + q(\bar q)\,.
\end{eqnarray}
In the next order we encounter the following processes. First we have
the bremsstrahlung correction to reaction (\ref{eq3.2})
\begin{eqnarray}
\label{eq3.4}
q + \bar q \rightarrow \gamma^* + g + g\,,
\end{eqnarray}
plus the two-loop correction to the Born process and the one-loop corrections
to process (\ref{eq3.2}). Next we have the bremsstrahlung correction to reaction
(\ref{eq3.3})
\begin{eqnarray}
\label{eq3.5}
g + q(\bar q) \rightarrow \gamma^* + q(\bar q) + g\,,
\end{eqnarray}
plus the one-loop correction to reaction (\ref{eq3.3}).
Subsequently we have three new reactions. The first involves a quark-anti-quark
reaction see Figs. \ref{fig1}, \ref{fig2}
\begin{eqnarray}
\label{eq3.6}
q + \bar q \rightarrow \gamma^* + q + \bar q \,,
\end{eqnarray}
and the second contains two (anti-) quarks in the initial state see Figs. 
\ref{fig2}, \ref{fig3} 
\begin{eqnarray}
\label{eq3.7}
q (\bar q) + q (\bar q) \rightarrow \gamma^* + q (\bar q) + q(\bar q)\,,
\end{eqnarray}
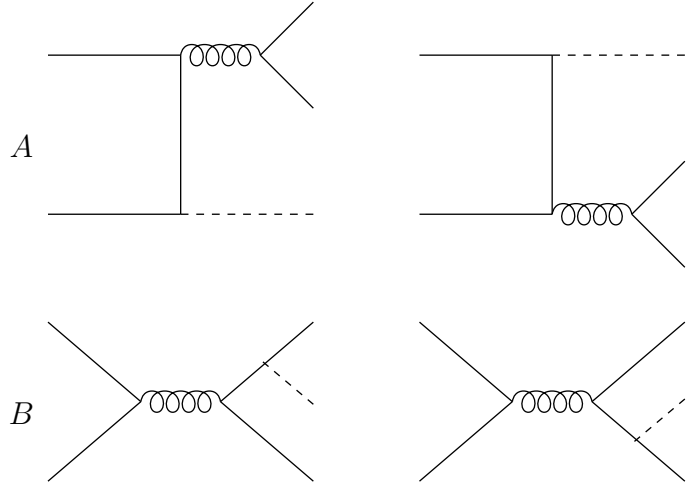
\begin{figure}
\begin{center}
\begin{picture}(260,95)(0,0)
\Text(0,50)[t]{$A$}
\Line(10,20)(60,20)
\Line(10,80)(60,80)
\Line(60,20)(60,80)
\Gluon(60,80)(90,80){4}{4}
\Line(90,80)(110,100)
\Line(90,80)(110,60)
\DashLine(60,20)(110,20){3}
\Line(150,20)(200,20)
\Line(150,80)(200,80)
\Line(200,20)(200,80)
\Gluon(200,20)(230,20){4}{4}
\Line(230,20)(250,0)
\Line(230,20)(250,40)
\DashLine(200,80)(250,80){3}

\end{picture}
\begin{picture}(260,100)(0,0)

\Text(0,50)[t]{$B$}

\Line(10,20)(45,50)
\Line(10,80)(45,50)
\Gluon(45,50)(75,50){4}{4}
\Line(75,50)(110,80)
\Line(75,50)(110,20)
\DashLine(91,65)(110,49){3}

\Line(150,20)(185,50)
\Line(150,80)(185,50)
\Gluon(185,50)(215,50){4}{4}
\Line(215,50)(250,20)
\Line(215,50)(250,80)
\DashLine(231,35)(250,51){3}
\end{picture}
\caption[]{Annihilation graphs contributing to the subprocess $q+\bar q
\rightarrow \gamma^* + q+\bar q$. }
\label{fig1}
\end{center}
\end{figure}
In the above reaction the quarks can be equal or unequal. Finally we
have a reaction with two gluons in the initial state
\begin{eqnarray}
\label{eq3.8}
g + g \rightarrow \gamma^* + q + \bar q\,.
\end{eqnarray}
All matrix elements are calculated in $n$-dimensions using the algebraic 
manipulation program FORM \cite{form}. The phase space integration and
the Feynman integration are explained in \cite{mama} and \cite{maha}.
The results can be expressed in the renormalization group coefficients
$\Delta P_{ij}$, $\beta_0$ and $w^{(i)}_{kl}$ as follows
\begin{eqnarray}
\label{eq3.9}
-\Delta \hat W_{q\bar q}^{\rm NS}&=&\delta(1-z)+\hat a_s \,S_\varepsilon\,
\left (\frac
{Q^2}{\mu^2} \right )^{\varepsilon/2}\,\Bigg [\frac{2}{\varepsilon}\,\Delta 
P^{(0)}_{qq} +w^{(1)}_{q\bar q}+2\,z^{(1)}_{qq}+
\nonumber\\[2ex]
&&+\varepsilon \,\left (\bar w^{(1)}_{q\bar q}
+2\,\bar z^{(1)}_{qq} \right )\Bigg ]+\hat a_s^2\,S_\varepsilon^2\,
\left (\frac{Q^2}{\mu^2} \right )^{\varepsilon}\,\Bigg [\frac{1}{\varepsilon^2}
\,\Big \{2\,\Delta P^{(0)}_{qq}\otimes \Delta P^{(0)}_{qq}
\nonumber\\[2ex]
&&-2\,\beta_0\,\Delta P^{(0)}_{qq}\Big \}
+\frac{1}{\varepsilon}\,\Big \{\Delta P^{(1),{\rm NS},+}_{qq}
+2\,\Delta P^{(0)}_{qq}\otimes w^{(1)}_{q\bar q}-2\,\beta_0\,w^{(1)}_{q\bar q}
\nonumber\\[2ex]
&&+4\,z^{(1)}_{qq} \otimes \Delta P^{(0)}_{qq}-2\,\beta_0\,z^{(1)}_{qq} \Big \}
+2\,\Delta P^{(0)}_{qq}\otimes \bar w^{(1)}_{q\bar q}-2\,\beta_0\,
\bar w^{(1)}_{q\bar q}
\nonumber\\[2ex]
&&+w^{(2),{\rm NS}}_{q\bar q}+4\,\bar z^{(1)}_{qq} \otimes \Delta P^{(0)}_{qq}
+2\,z^{(1)}_{qq}\otimes w^{(1)}_{q\bar q}
+z^{(1)}_{qq}\otimes z^{(1)}_{qq}
\nonumber\\[2ex]
&& -4\,\beta_0\,\bar z^{(1)}_{qq}+2\,z^{(2),{\rm NS},+}_{qq}\Bigg ]\,,
\end{eqnarray}
\begin{figure}
\begin{center}
  \begin{picture}(260,90)(0,0)

\Text(0,50)[t]{$C$}

\Line(10,20)(110,20)
\Line(10,70)(110,70)
\Gluon(60,20)(60,70){4}{6}
\DashLine(35,70)(70,90){3}

\Line(150,20)(250,20)
\Line(150,70)(250,70)
\Gluon(200,20)(200,70){4}{5}
\DashLine(225,70)(260,90){3}

\Text(0,70)[t]{$1$}
\Text(0,20)[t]{$2$}
\Text(140,70)[t]{$1$}
\Text(140,20)[t]{$2$}

\Text(120,70)[t]{$3$}
\Text(120,20)[t]{$4$}
\Text(260,70)[t]{$3$}
\Text(260,20)[t]{$4$}

\end{picture}

\vspace*{3mm}
\begin{picture}(260,100)(0,0)

\Text(0,50)[t]{$D$}

\Line(10,20)(110,20)
\Line(10,70)(110,70)
\Gluon(60,20)(60,70){4}{6}
\DashLine(35,20)(70,0){3}

\Line(150,20)(250,20)
\Line(150,70)(250,70)
\Gluon(200,20)(200,70){4}{5}
\DashLine(225,20)(260,0){3}

\Text(0,70)[t]{$1$}
\Text(0,20)[t]{$2$}
\Text(140,70)[t]{$1$}
\Text(140,20)[t]{$2$}

\Text(120,70)[t]{$3$}
\Text(120,20)[t]{$4$}
\Text(260,70)[t]{$3$}
\Text(260,20)[t]{$4$}

\end{picture}

\caption[]{Gluon exchange graphs contributing to the subprocess $q+\bar q
\rightarrow \gamma^* +q+\bar q$ and $q + q \rightarrow \gamma^* +q + q$. }
\label{fig2}
\end{center}
\end{figure}
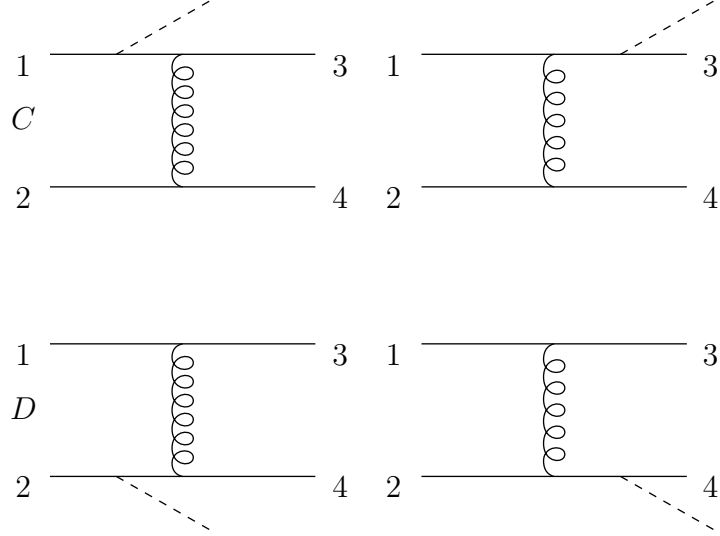
\begin{eqnarray}
\label{eq3.10}
\Delta \hat W_{qg}&=&-\hat a_s\,S_\varepsilon\,\left (\frac{Q^2}{\mu^2}\right )^
{\varepsilon/2}\,\Bigg [\frac{1}{2\varepsilon}\,\Delta P_{qg}^{(0)}+w^{(1)}_{qg}
+\varepsilon\,\bar w^{(1)}_{qg}\Bigg ]-\hat a_s^2\,S_\varepsilon^2\,
\left (\frac{Q^2}{\mu^2}\right )^\varepsilon
\nonumber\\[2ex]
&&\times \Bigg [\frac{1}{\varepsilon^2}\,
\Big \{ \frac{3}{4}\,\Delta P^{(0)}_{qq}\otimes \Delta P^{(0)}_{qg}
+\frac{1}{4}\,\Delta P^{(0)}_{gg}
\otimes \Delta P^{(0)}_{qg}-\frac{1}{2}\,\beta_0\,\Delta P^{(0)}_{qg}\Big \}
\nonumber\\[2ex]
&&+\frac{1}{\varepsilon}\,\Big \{\frac{1}{4}\,\Delta P^{(1)}_{qg}
-2\,\beta_0\,w^{(1)}_{qg} +\frac{1}{2}\,\Delta P^{(0)}_{qg}\otimes
w^{(1)}_{q\bar q}
+\Big (\Delta P^{(0)}_{qq}+\Delta P^{(0)}_{gg}\Big )
\nonumber\\[2ex]
&&\otimes w^{(1)}_{qg}+\frac{3}{4}\,z^{(1)}_{qq}\otimes \Delta P^{(0)}_{qg}
\Big \}
-2\,\beta_0\,\bar w^{(1)}_{qg}+\frac{1}{2}\,\Delta P^{(0)}_{qg}\otimes
\bar w^{(1)}_{q\bar q}+\Big (\Delta P^{(0)}_{qq}
\nonumber\\[2ex]
&&+\Delta P^{(0)}_{gg}\Big )\otimes
\bar w^{(1)}_{qg}+w^{(2)}_{qg}
+\frac{1}{2}\,\bar z^{(1)}_{qq}\otimes \Delta P^{(0)}_{qg}
+z^{(1)}_{qq}\otimes w^{(1)}_{qg}\Bigg ]\,.
\end{eqnarray}
The $q\bar q$ reaction (see Figs. \ref{fig1},\ref{fig2}) yields
\begin{eqnarray}
\label{eq3.11}
\Delta \hat W_{q\bar q}^{\rm PS}&=&-\hat a_s^2\,S_\varepsilon^2\,\left (
\frac{Q^2}{\mu^2} \right )
^\varepsilon\,\Bigg [\frac{1}{2\varepsilon^2}\,\Delta P^{(0)}_{qg}\otimes 
\Delta P^{(0)}_{gq}
+\frac{1}{\varepsilon}\,\Big \{ \frac{1}{2}\,\Delta P^{(1),{\rm PS}}_{qq}
\nonumber\\[2ex]
&&+2\, \Delta P^{(0)}_{gq}\otimes w^{(0)}_{qg}\Big \}
+2\,\Delta P^{(0)}_{gq}\otimes \bar w^{(1)}_{qg}
+w^{(2),{\rm PS}}_{q\bar q}+2\,z^{(2),{\rm PS}}_{qq}\Bigg ] \,.
\nonumber\\[2ex]
\end{eqnarray}
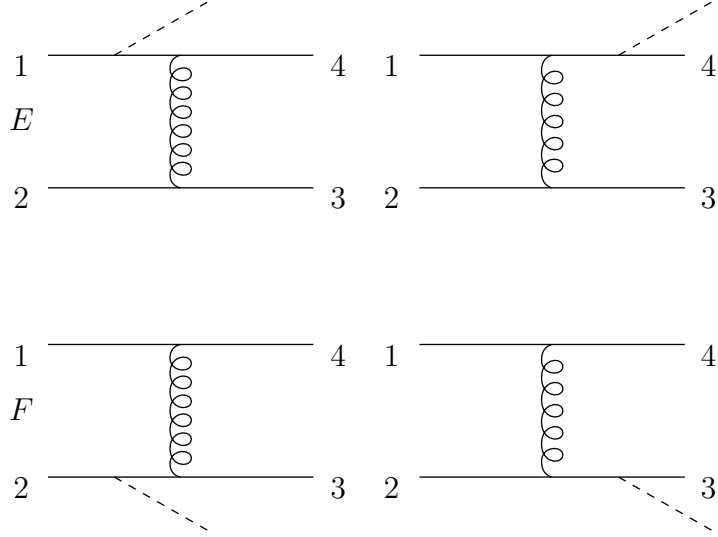
\begin{figure}
\begin{center}
  \begin{picture}(260,85)(0,0)

\Text(0,50)[t]{$E$}

\Line(10,20)(110,20)
\Line(10,70)(110,70)
\Gluon(60,20)(60,70){4}{6}
\DashLine(35,70)(70,90){3}

\Line(150,20)(250,20)
\Line(150,70)(250,70)
\Gluon(200,20)(200,70){4}{5}
\DashLine(225,70)(260,90){3}

\Text(0,70)[t]{$1$}
\Text(0,20)[t]{$2$}
\Text(140,70)[t]{$1$}
\Text(140,20)[t]{$2$}

\Text(120,70)[t]{$4$}
\Text(120,20)[t]{$3$}
\Text(260,70)[t]{$4$}
\Text(260,20)[t]{$3$}

\end{picture}

\vspace*{3mm}
\begin{picture}(260,100)(0,0)

\Text(0,50)[t]{$F$}

\Line(10,20)(110,20)
\Line(10,70)(110,70)
\Gluon(60,20)(60,70){4}{6}
\DashLine(35,20)(70,0){3}

\Line(150,20)(250,20)
\Line(150,70)(250,70)
\Gluon(200,20)(200,70){4}{5}
\DashLine(225,20)(260,0){3}

\Text(0,70)[t]{$1$}
\Text(0,20)[t]{$2$}
\Text(140,70)[t]{$1$}
\Text(140,20)[t]{$2$}

\Text(120,70)[t]{$4$}
\Text(120,20)[t]{$3$}
\Text(260,70)[t]{$4$}
\Text(260,20)[t]{$3$}

\end{picture}
\caption[]{Gluon exchange graphs contributing to the subprocess
$q + q \rightarrow \gamma^* +q + q$ with identical quarks in the initial
and/or final state. }
\label{fig3}
\end{center}
\end{figure}
The $q\bar q$ can be split into $q_1\bar q_2$ ($q_1 \not =q_2$)
and a $q\bar q$ reaction. The latter does include Fig. \ref{fig1}B.
\begin{eqnarray}
\label{eq3.12}
w^{(2),{\rm PS}}_{q_1\bar q_2}=w^{(2)}_{CC}+w^{(2)}_{DD}+w^{(2)}_{CD}\,,
\end{eqnarray}
\begin{eqnarray}
\label{eq3.13}
w^{(2),{\rm PS}}_{q\bar q}=w^{(2)}_{BB}+w^{(2)}_{BC}+w^{(2)}_{BD}+w^{(2)}_{CC}
+w^{(2)}_{DD} +w^{(2)}_{CD}\,.
\end{eqnarray}
Processes with unequal quarks in the final state 
(see Fig. \ref{fig2}) yields
\begin{eqnarray}
\label{eq3.14}
\Delta \hat W_{q_1q_2}&=&-\hat a_s^2\,S_\varepsilon^2\,\left (\frac{Q^2}{\mu^2}
\right )
^\varepsilon\,\Bigg [\frac{1}{2\varepsilon^2}\,\Delta P^{(0)}_{qg}\otimes 
\Delta P^{(0)}_{gq} +\frac{1}{\varepsilon}\,\Big \{ \frac{1}{2}\,\Delta 
P^{(1),{\rm PS}}_{qq}
\nonumber\\[2ex]
&&+2\,\Delta P^{(0)}_{gq}\otimes w^{(0)}_{qg}\Big \}
+2\,\Delta P^{(0)}_{gq}\otimes \bar w^{(1)}_{qg}
+w^{(2)}_{q_1q_2}+2\,z^{(2),{\rm PS}}_{qq}\Bigg ]\,,
\end{eqnarray}
\begin{eqnarray}
\label{eq3.15}
w^{(2)}_{q_1q_2}=w^{(2)}_{CC}+w^{(2)}_{DD}+w^{(2)}_{CD}\,.
\end{eqnarray}
Processes with equal quarks in the final state (see Figs. \ref{fig2}, 
\ref{fig3}) yields
\begin{eqnarray}
\label{eq3.16}
\Delta \hat W_{qq}&=&-\hat a_s^2\,S_\varepsilon^2\,\left (\frac{Q^2}{\mu^2}
\right )
^\varepsilon\,\Bigg [\frac{1}{2\varepsilon^2}\,\Delta P^{(0)}_{qg}\otimes 
\Delta P^{(0)}_{gq}
+\frac{1}{\varepsilon}\,\Big \{ \frac{1}{2}\,\Delta P^{(1),{\rm PS}}_{qq}-
\Delta P^{(1),{\rm NS},-}_{qq}
\nonumber\\[2ex]
&& +2\,\Delta P^{(0)}_{gq}\otimes w^{(0)}_{qg}\Big \}
+2\,\Delta P^{(0)}_{gq}\otimes \bar w^{(1)}_{qg} +w^{(2)}_{qq}+
2\,z^{(2),{\rm PS}}_{qq} 
\nonumber\\[2ex]
&&-2\,z^{(2),{\rm NS},-}_{qq}\Bigg ]\,,
\end{eqnarray}
\begin{eqnarray}
\label{eq3.17}
w^{(2)}_{qq}&=&w^{(2)}_{CC}+w^{(2)}_{DD}+w^{(2)}_{CD}+w^{(2)}_{FF}+w^{(2)}_{EE}
+w^{(2)}_{FE}+w^{(2)}_{CF}+w^{(2)}_{CE}+w^{(2)}_{DF}
\nonumber\\[2ex]
&&+w^{(2)}_{DE}\,.
\end{eqnarray}
Finally we have the gluon-gluon scattering process
\begin{eqnarray}
\label{eq3.18}
\Delta \hat W_{gg}&=&-\hat a_s^2\,S_\varepsilon^2\,\left (\frac{Q^2}{\mu^2}
\right )
^\varepsilon\,\Bigg [\frac{1}{2\varepsilon^2}\,\Delta P^{(0)}_{qg}\otimes 
\Delta P^{(0)}_{qg}
+\frac{2}{\varepsilon}\,\Delta P^{(0)}_{qg}\otimes w^{(1)}_{qg} 
\nonumber\\[2ex]
&&+2\,\Delta P^{(0)}_{qg} \otimes \bar w^{(1)}_{qg} +w^{(2)}_{gg}\Bigg ]\,,
\end{eqnarray}
where $\otimes$ denotes the convolution symbol defined by 
\begin{eqnarray}
\label{eq3.19}
f\otimes g(z)=\int_0^1dz_1\,\int_0^1dz_2\,\delta(z-z_1\,z_2)\,f(z_1)\,g(z_2)\,.
\end{eqnarray}
The expressions follow from the renormalization group equations. They are
constructed in such a way that they become finite after coupling constant 
renormalization and mass factorization are carried out. 
The $\beta_0$ appears as the lowest order coefficient in the beta-function.
The splitting
function $\Delta P_{ij}(z)$ and the coefficients $w_{ij}^{(i)}(z)$ with 
$z=Q^2/s$ also occur in the coefficient functions given below except for the 
NLO terms $\bar w_{ij}^{(1)}$, which are proportional to $\varepsilon$. 
They are given by
\begin{eqnarray}
\label{eq3.20}
\bar w^{(1)}_{q\bar q}&=&C_F\,\Bigg [8\,{\cal D}_2(x)-6\,\zeta(2)\,
{\cal D}_0(x)+(1+x)\,
\Big (-4\,\ln^2(1-x)+3\,\zeta(2)\Big )
\nonumber\\[2ex]
&&+\frac{1+x^2}{1-x}\,(-4\,\ln x\,\ln (1-x)
+\ln^2 x)+4\,(1-x)+\delta(1-x)\,\Big (16
\nonumber\\[2ex]
&&-\frac{21}{2}\,\zeta(2)\Big )\Bigg ]\,,
\end{eqnarray}
\begin{eqnarray}
\label{eq3.21}
\bar w^{(1)}_{qg}&=&T_f\,\Bigg [\frac{1}{2}\,(2\,x-1)\,\Big (\ln^2\left (
\frac{(1-x)^2}{x}\right ) -3\,\zeta(2)\Big )+(\frac{5}{2}-x-\frac{3}{2}\,x^2)\,
\nonumber\\[2ex]
&&\times \ln \left (\frac{(1-x)^2}{x} \right)-5+4\,x+x^2\Bigg ]\,,
\end{eqnarray}
where ${\cal D}_i$ denotes the distribution
\begin{eqnarray}
\label{eq3.22}
{\cal D}_i=\left (\frac{\ln^i(1-z)}{1-z}\right )_+\,.
\end{eqnarray}
To render the partonic cross sections finite one has first to perform 
coupling constant renormalization. This is done by replacing the bare coupling
constant by the renormalized one i.e.
\begin{eqnarray}
\label{eq3.23}
\hat a_s=a_s(\mu^2)\,\Bigg [1+a_s(\mu^2)\,S_\varepsilon\,
\frac{2}{\varepsilon}\,\beta_0 +\cdots \Bigg ]\,.
\end{eqnarray}
To remove the evanescent counter terms we perform the following operation.
We multiply each quark line with the inverse of $Z_{qq}^{5}$ \cite{neerv}.
This means that $q\bar q$, $qq$ and $q_1q_2$ are multiplied by 
$(Z_{qq}^{5})^{-2}$. The $qg$ is multiplied by $(Z_{qq}^{5})^{-1}$
and the $gg$ subprocess gets no factor. The remaining divergences, which
are of collinear origin, are removed by mass factorization
\begin{eqnarray}
\label{eq3.24}
\left (Z_{qq}^{5}\right )^{-\alpha}\,\Delta \hat W_{ij}
\left (\frac{1}{\varepsilon},
\frac{Q^2}{\mu^2}\right )
=\sum_{k,l=q,\bar q,g}\,\Gamma_{ki}\left (\frac{1}{\varepsilon}\right )
\otimes \Gamma_{lj}\left (\frac{1}{\varepsilon}\right )\otimes \Delta_{kl}
\left (\frac{Q^2}{\mu^2}\right )\,,
\end{eqnarray}
where $\alpha$ is the number of (anti)-quarks in the initial state. Further
$\Gamma_{ki}(z)$ denote the kernels containing the splitting function
which multiply the collinear divergences represented by the pole terms
$1/\varepsilon^k$. For the different subprocesses the mass factorization
relations in Eq.(3.24) become equal to
\begin{eqnarray}
\label{3.25}
\left (Z_{qq}^{5}\right )^{-2}\,\Delta \hat W_{q\bar q}^{\rm NS}=
\Gamma_{qq}^{\rm NS}
\otimes \Gamma_{\bar q\bar q}^{\rm NS} \otimes \Delta_{q\bar q}^{\rm NS}\,,
\end{eqnarray}
\begin{eqnarray}
\label{eq3.26}
\left (Z_{qq}^{5}\right )^{-1}\,\Delta \hat W_{qg}=\Gamma_{qq}\otimes 
\Gamma_{\bar qg}\otimes \Delta_{q\bar q}
+\Gamma_{qq}\otimes \Gamma_{gg}\otimes \Delta_{qg}\,,
\end{eqnarray}
\begin{eqnarray}
\label{eq3.27}
\left (Z_{qq}^{5}\right )^{-2}\,\Delta \hat W_{q\bar q}=\Gamma_{qq}\otimes 
\Gamma_{\bar q
\bar q}\otimes \Delta_{q \bar q}+\Gamma_{g\bar q}\otimes \Gamma_{qq}
\otimes \Delta_{gq}+\Gamma_{gq}\otimes \Gamma_{\bar q\bar q}\otimes 
\Delta_{g\bar q}\,,
\end{eqnarray}
\begin{eqnarray}
\label{eq3.28}
\left (Z_{qq}^{5}\right )^{-2}\,\Delta \hat W_{q_1q_2}&=&\Gamma_{q_1q_1}\otimes
\Gamma_{q_1q_2} \otimes \Delta_{qq}+\Gamma_{gq_1}\otimes \Gamma_{q_2q_2}
\otimes \Delta_{gq_2}
\nonumber\\[2ex]
&&+\Gamma_{gq_2}\otimes \Gamma_{q_1q_1}\otimes \Delta_{gq_1}\,,
\end{eqnarray}
\begin{eqnarray}
\label{eq3.29}
\left (Z_{qq}^{5}\right )^{-2}\,\Delta \hat W_{qq}&=&\Gamma_{qq}\otimes 
\Gamma_{\bar q
q}\otimes \Delta_{q \bar q}+\Gamma_{\bar qq}\otimes \Gamma_{qq}\otimes 
\Delta_{\bar qq}+ \Gamma_{gq}\otimes \Gamma_{qq}
\otimes \Delta_{gq}
\nonumber\\[2ex]
&&+\Gamma_{qq}\otimes \Gamma_{gq}\otimes 
\Delta_{qg}+\Gamma_{qq}\otimes \Gamma_{qq}\otimes \Delta_{qq}\,,
\end{eqnarray}
\begin{eqnarray}
\label{eq3.30}
\Delta \hat W_{gg}&=&2\,\Gamma_{qg}\otimes \Gamma_{\bar qg}\otimes 
\Delta_{q \bar q}
+2\,\Gamma_{\bar qg}\otimes \Gamma_{gg}\otimes 
\Delta_{qg}+ 2\,\Gamma_{\bar qg}\otimes \Gamma_{gg}
\otimes \Delta_{\bar qg}
\nonumber\\[2ex]
&&+\Gamma_{gg}\otimes \Gamma_{gg}\otimes \Delta_{gg}\,.
\end{eqnarray}
Since we need the finite expressions up to order $\alpha_s^2$ it is sufficient
to expand the kernels $\Gamma_{ki}$ up to the following order in the 
renormalized coupling constant
\begin{eqnarray}
\label{eq3.31}
\Gamma_{qq}^{\rm NS}&=&\Gamma_{\bar q\bar q}^{\rm NS}=\delta(1-z)
+\hat a_s\,S_\varepsilon
\,\Bigg [\frac{1}{\varepsilon}\,\Delta P_{qq}^{(0)}\Bigg ]
+\hat a_s^2\,S_\varepsilon^2\,\Bigg [\frac{1}{2\varepsilon^2}
\Big (\Delta P_{qq}^{(0)} \otimes \Delta P_{qq}^{(0)}
\nonumber\\[2ex]
&&-2\,\beta_0\,\Delta P_{qq}^{(0)}\Big )+\frac{1}{2\varepsilon}
\,\Delta P_{qq}^{(1),{\rm NS},+}\Bigg ]\,,
\end{eqnarray}
\begin{eqnarray}
\label{eq3.32}
\Gamma_{qg}&=&\Gamma_{\bar qg}=\hat a_s\,S_\varepsilon\,\Bigg [
\frac{1}{2\varepsilon}\,\Delta P_{qg}^{(0)}\Bigg ]+\hat a_s^2\,S_\varepsilon^2\,
\Bigg [\frac{1}{4\varepsilon^2}\Big (\Delta P_{qg}^{(0)}\otimes \Delta 
P_{gg}^{(0)} 
\nonumber\\[2ex]
&&+\Delta P_{qq}^{(0)}\otimes \Delta P_{qg}^{(0)}-2\,\beta_0\,
\Delta P_{qg}^{(0)}\Big )+\frac{1}{4\varepsilon}\,\Big (\Delta P_{qg}^{(1)}
\nonumber\\[2ex]
&&+\Delta P^{(0)}_{qg} \otimes z^{(1)}_{qq}\Big )\Bigg ]\,,
\end{eqnarray}
\begin{eqnarray}
\label{eq3.33}
\Gamma_{qq}^{\rm PS}=\Gamma_{\bar q\bar q}^{\rm PS}=\Gamma_{q\bar q}^{\rm PS}=
\Gamma_{\bar qq}^{\rm PS}=
\hat a_s^2\,S_\varepsilon^2\,\Bigg [
\frac{1}{4\varepsilon^2}\,\Delta P_{qg}^{(0)}\otimes \Delta P_{gq}^{(0)}+
\frac{1}{4\varepsilon}\,\Delta P_{qq}^{(1),{\rm PS}}\Bigg ]\,,
\end{eqnarray}
\begin{eqnarray}
\label{eq3.34}
\Gamma_{q\bar q}^{\rm NS}=\Gamma_{\bar qq}^{\rm NS}=
-\hat a_s^2\,S_\varepsilon^2\,\Bigg [
\frac{1}{2\varepsilon}\,\Delta P_{qq}^{(1),{\rm NS},-}\Bigg ]\,,
\end{eqnarray}
\begin{eqnarray}
\label{eq3.35}
\Gamma_{gq}=\Gamma_{g\bar q}=\hat a_s\,S_\varepsilon\,\Bigg [
\frac{1}{\varepsilon}\,\Delta P_{gq}^{(0)}\Bigg ]\,,
\end{eqnarray}
\begin{eqnarray}
\label{eq3.36}
\Gamma_{gg}=\delta(1-z)+\hat a_s\,S_\varepsilon\,\Bigg [\frac{1}{\varepsilon}\,
\Delta P_{gg}^{(0)}\Bigg ]\,.
\end{eqnarray}
After renormalization and mass factorization the coefficient functions have the
following algebraic form
\begin{eqnarray}
\label{eq3.37}
\Delta_{q\bar q}^{\rm NS}&=&-\delta(1-z)-a_s(\mu^2)\,\Bigg [\Delta P^{(0)}_{qq}
\,\ln \frac{Q^2}{\mu^2}+w^{(1)}_{q\bar q}\Bigg ]-a_s^2(\mu^2)\,\Bigg [  
\nonumber\\[2ex]
&&\Big \{ \frac{1}{2}\,\Delta P^{(0)}_{qq}\otimes \Delta P^{(0)}_{qq}
-\frac{1}{2}\,
\beta_0\,\Delta P^{(0)}_{qq} \Big \}\,\ln^2\frac{Q^2}{\mu^2}+\Big \{\Delta
P^{(1),{\rm NS},+} 
\nonumber\\[2ex]
&&-\beta_0\,w^{(1)}_{q\bar q}+\Delta P^{(0)}_{qq} \otimes w^{(1)}_{q\bar q} 
\Big \} \ln \frac{Q^2}{\mu^2} +w^{(2),{\rm NS}}_{q\bar q}\Bigg ]\,,
\end{eqnarray}
\begin{eqnarray}
\label{3.38}
\Delta_{qg}&=&-a_s(\mu^2)\,\Bigg [\frac{1}{4}\,\Delta P^{(0)}_{qg}\,\ln 
\frac{Q^2}{\mu^2}
+w^{(1)}_{qg}\Bigg ]-a_s^2(\mu^2)\,\Bigg [\Big \{\frac{1}{16}\,
\Delta P^{(0)}_{qg}\otimes \Big (3\,\Delta P^{(0)}_{qq}
\nonumber\\[2ex]
&&+\Delta P^{(0)}_{gg}\Big )-\frac{1}{8}\,\beta_0\,\Delta P^{(0)}_{qg}
\Big \}\,\ln^2\frac{Q^2}{\mu^2} +\Big \{\frac{1}{4}\Delta P^{(1)}_{qg}
-\beta_0\,w^{(1)}_{qg}+\frac{1}{2}\,w^{(1)}_{qg}
\nonumber\\[2ex]
&&\otimes \Big (\Delta P^{(0)}_{qq}
+\Delta P^{(0)}_{gg}\Big )+\frac{1}{4}\,\Delta P^{(0)}_{qg}
\otimes w^{(1)}_{q\bar q}+\frac{1}{4}\,\Delta P^{(0)}_{qg}\otimes z^{(1)}_{qq}
\Big \}\,\ln \frac{Q^2}{\mu^2} 
\nonumber\\[2ex]
&&+ w^{(2)}_{qg}\Bigg ]\,,
\end{eqnarray}
\begin{eqnarray}
\label{eq3.39}
\Delta_{q\bar q}^{\rm PS}&=&-a_s^2(\mu^2)\,\Bigg [\Big \{\frac{1}{8}
\Delta P^{(0)}_{qg}
\otimes \Delta P^{(0)}_{gq} \Big \}\, \ln^2 \frac{Q^2}{\mu^2}+\Big \{
\frac{1}{2}\,\Delta P^{(1),{\rm PS}}_{qq}
\nonumber\\[2ex]
&&+\Delta P^{(0)}_{gq} \otimes w^{(1)}_{qg}\Big \}
 \ln \frac{Q^2}{\mu^2}+w^{(2),{\rm PS}}_{q\bar q}\Bigg ]\,,
\end{eqnarray}
\begin{eqnarray}
\label{eq3.40}
\Delta_{q_1q_2}&=&-a_s^2(\mu^2)\,\Bigg [\Big \{\frac{1}{8}\Delta P^{(0)}_{qg}
\otimes \Delta P^{(0)}_{gq} \Big \}\, \ln^2 \frac{Q^2}{\mu^2}+\Big \{
\frac{1}{2}\,\Delta P^{(1),{\rm PS}}_{qq}
\nonumber\\[2ex]
&&+\Delta P^{(0)}_{gq}\otimes w^{(1)}_{qg}\Big \}
 \ln \frac{Q^2}{\mu^2}+w^{(2)}_{q_1q_2}\Bigg ]\,,
\end{eqnarray}
\begin{eqnarray}
\label{eq3.41}
\Delta_{qq}&=&-a_s^2(\mu^2)\,\Bigg [\Big \{\frac{1}{8}\Delta P^{(0)}_{qg}
\otimes \Delta P^{(0)}_{gq} \Big \}\, \ln^2 \frac{Q^2}{\mu^2}+\Big \{
\frac{1}{2}\,\Delta P^{(1),{\rm PS}}_{qq}-\Delta P^{(1),{\rm NS},-}_{q\bar q}
\nonumber\\[2ex]
&&+ \Delta P^{(0)}_{gq}\otimes w^{(1)}_{qg}\Big \}
\ln \frac{Q^2}{\mu^2}+w_{qq}^{(2)}\Bigg ]\,,
\end{eqnarray} 
\begin{eqnarray}
\label{eq3.42}
\Delta_{gg}&=&-a_s^2(\mu^2)\,\Bigg [\Big \{\frac{1}{8}\,\Delta P_{qg}^{(0)}
\otimes
\Delta P_{qg}^{(0)}\Big \}\,\ln^2 \frac{Q^2}{\mu^2}+
\Big \{\Delta P_{qg}^{(0)}\otimes w_{qg}^{(1)} \Big \}\,\ln \frac{Q^2}{\mu^2}
\nonumber\\[2ex]
&& +w_{gg}^{(2)}\Bigg ]\,.
\end{eqnarray}
In Appendix A we give the explicit expressions for the coefficient
functions so that one can determine the coefficients $\Delta P_{ij}^{(k)}$
and $w_{ij}^{(k)}$. One of the features is that the non-singlet part of
the structure function satisfies the relation
\begin{eqnarray}
\label{eq3.43}
\left (Z_{qq}^{5,{\rm NS},+}\right )^{-2}\,\Delta \hat W_{q\bar q}^{\rm NS}=
-\hat W_{q\bar q}^{\rm NS}\,.
\end{eqnarray}
For the O($\alpha_s^2$) correction this holds for 
Eqs. (\ref{eqA.1})-(\ref{eqA.13}).
In the presentation of the coefficient functions
above we have put the renormalization scale $\mu_r$ equal to the mass 
factorization scale $\mu$. If one wants to distinguish between both scales one 
can make the simple substitution
\begin{eqnarray}
\label{eq3.44}
\alpha_s(\mu^2)=\alpha_s(\mu_r^2)\,\Bigg [1+\frac{\alpha_s(\mu_r^2)}{4\pi}
\,\beta_0\,\ln \frac{\mu_r^2}{\mu^2}\Bigg ]\,.
\end{eqnarray}

\mysection{$d\Delta \sigma/dQ$ for the process
 $p + p\rightarrow \gamma^* +'X'$}
In this section we will present total cross sections (see Eq. (\ref{eq2.2}))
for polarized Drell-Yan production in proton-proton collisions at the RHIC
and make a comparison with similar results in previous work. The cross 
section can be rewritten as
\begin{eqnarray}
\label{eq4.1}
\frac{d\Delta \sigma}{dQ}=\frac{8\,\pi\,\alpha^2}{3\,N\,S\,Q}\,
\sum_{i,j=q,\bar q,g}\,\int_{\tau}^1\frac{dy}{y}\,
\Phi_{ij}(y,\mu^2)\,\Delta_{ij}\left (\frac{\tau}{y},\frac{Q^2}{\mu^2}\right )
\,,
\end{eqnarray}
where $\tau=Q^2/S$ and $\Phi_{ij}$ is the parton-parton flux defined by
\begin{eqnarray}
\label{eq4.2}
\Phi_{ij}(y,\mu^2)=\int_y^1\frac{du}{u}\,\Delta f_i(u,\mu^2)\,\Delta f_j
\left (\frac{y}{u}, \mu^2\right )\,.
\end{eqnarray}
In particular we study the dependence of the cross section on the input 
parameters 
such as the renormalization/factorization scale $\mu$, the virtuality of the 
photon $Q$ and the input parton densities. At this moment LO and NLO
polarized parton densities are available but NNLO are not. Therefore an exact
NNLO polarized cross section cannot be determined yet. Even an approximated 
polarized cross section cannot be given because a finite moment analysis
of the anomalous dimensions is not yet available. Only the non-singlet
anomalous dimension is known at least up to $N=14$ \cite{rv}. Because of a lack
of data the polarized parton density sets show a larger variation than in 
the unpolarized density sets. This in particular
holds for the sea-quark and the gluon densities. For this reason polarized
Drell-Yan production is very useful. 
The sea-quark contribution is dominant for the totally integrated cross
section in proton-proton collisions  while the gluon contribution dominates
the differential distribution w.r.t. $p_T$ for $p_T>Q/2$.
We shall use BB set 1 (BB1) for our plots \cite{blbo} in the whole paper
except for the longitudinal asymmetry where we shall choose also other 
density sets like BB set 2 (BB2) \cite{blbo}, the
GRSV01 (standard scenario) and GRSV01 (valence
scenario) \cite{grsv}. The gluon in BB1 is larger than in BB2. In its turn
the gluon in the standard scenario is smaller than in BB2 but larger than in 
the valence scenario. For all these sets LO and NLO versions exist.
It is clear that in the case of LO we will use the one-loop parametrization
for the running coupling constant and for NLO the two-loop corrected running
coupling constant. However for an estimate of the NNLO corrections
we will only choose NLO polarized parton densities with the two-loop running
coupling constant.
\begin{table}
\begin{center}
\begin{tabular}{|c|c|c|}\hline
GRSV01 (LO, standard scenario)& $\Lambda_4^{\rm LO}=175~{\rm MeV}$  &
$\alpha_s^{\rm LO}(M_Z)=0.121$  \\
GRSV01 (NLO, standard scenario)  & $\Lambda_4^{\rm NLO}=257~{\rm MeV}$ &
$\alpha_s^{\rm NLO}(M_Z)=0.109$       \\
GRSV01 (LO, valence scenario)     & $\Lambda_4^{\rm LO}=175~{\rm MeV}$   &
$\alpha_s^{\rm LO}(M_Z)=0.121$       \\
GRSV01 (NLO, valence scenario)    & $\Lambda_4^{\rm NLO}=257~{\rm MeV}$  &
$\alpha_s^{\rm NLO}(M_Z)=0.109$    \\
BB (LO, scenario 1) & $\Lambda_4^{\rm LO}=203~{\rm MeV}$   &
$\alpha_s^{\rm LO}(M_Z)=0.123$   \\
BB (NLO, scenario 1) & $\Lambda_4^{\rm NLO}=235~{\rm MeV}$  &
$\alpha_s^{\rm NLO}(M_Z)=0.107$       \\
BB (LO, scenario 2) & $\Lambda_4^{\rm LO}=195~{\rm MeV}$  &
$\alpha_s^{\rm LO}(M_Z)=0.123$       \\
BB (NLO, scenario 2) & $\Lambda_4^{\rm NLO}=240~{\rm MeV}$ &
$\alpha_s^{\rm NLO}(M_Z)=0.107$ \\
MRST02(LO, lo2002.dat) & $\Lambda_4^{\rm LO}=220~{\rm MeV}$ &
$\alpha_s^{\rm LO}(M_Z)=0.125$ \\
MRST01(NLO, alf119.dat) & $\Lambda_4^{\rm NLO}=323~{\rm MeV}$ &
$\alpha_s^{\rm NLO}(M_Z)=0.113$ \\
MRST02(NNLO, vnval1155.dat)& $\Lambda_4^{\rm NLO}=235~{\rm MeV}$ &
$\alpha_s^{\rm NNLO}(M_Z)=0.109$ \\
\hline
\end{tabular}
\end{center}
\caption{Polarized and unpolarized parton density sets with the values for 
the QCD scale $\Lambda_4$ and the running coupling $\alpha_s(M_Z)$.}
\label{table1}
\end{table}
The details are given in Table \ref{table1}. Further we put $n_f=4$
in the coefficient functions and the running coupling constants and
the densities are only presented for the u,d,s and g partons. Finally we also
make a comparison with unpolarized Drell-Yan production. For this cross section
we adopt for LO the MRST02 \cite{maro1} densities, for NLO the
MRST01 \cite{maro2} densities and for NNLO the 
approximate MRST02 \cite{maro1} densities, with one, two and three
loop running coupling constants respectively.

In Fig. 4 we have plotted the polarized Drell-Yan cross section in Eq. 
(\ref{eq4.1}) up to NLO. It is clear that the $q\bar q$ contribution 
dominates the $qg$ contribution. The former is negative above $Q=$ 25 GeV, 
while the latter is already negative in a region above $Q$=13 GeV. 
This implies that
the total NLO contribution is actually negative above $Q$=27 GeV where 
it is very small.  For this reason we have plotted the absolute values 
of the contributions in Fig. 4.

The relative sizes are not changed if we go to NNLO as demonstrated in Fig.5.
Again the $q\bar q$ contribution is negative above $Q$= 25 GeV and the
$qg$ contribution is negative between $Q$= 4 GeV and $Q$= 13 GeV.
Here the $qq$ and $gg$ channels appear for the first time and they
are both negative.  However their contribution is negligible
compared with the one given by $qg$ and certainly by the $q\bar q$ result.
Again we have plotted the absolute values of the contributions.
We conclude that via the $q\bar q$ subprocess the sea-quark contribution
dominates the whole cross section in proton-proton collisions 
irrespective of the value taken by the gluon density.

In Fig. 6 we have plotted the absolute values of the LO, NLO and NNLO 
polarized cross sections for $2<Q<30$ GeV. 
Notice that we are in the range of small $Q$-values
so the $K$-factors are pretty large. We have defined the $K$-factors as
\begin{eqnarray}
\label{eq4.3}
K^{\rm NLO}=\frac{\Delta \sigma^{\rm NLO}}{\Delta \sigma^{\rm LO}}\,,
\qquad
K^{\rm NNLO}=\frac{\Delta \sigma^{\rm NNLO}}{\Delta \sigma^{\rm LO}}\,,
\end{eqnarray}
and plotted them in Fig. 7. At $Q=7~{\rm GeV}$ the K-factors reach a minimum,
$K^{\rm NLO}\sim 1.2$ and $K^{\rm NNLO}\sim 1.3$, and at larger $Q$ they
rise again. The rapid rise
above $Q$= 18 GeV is due to the change in sign of the LO process
near $Q$=25 GeV where the ratio is infinite. However the magnitude of the
polarized cross section is extremely small in this region.

Next we turn to the variation of the cross section w.r.t. scale $\mu$.
In Figs. 8a, 8b and 8c we show the LO, NLO and NNLO polarized cross sections
at the scales $\mu=Q/2$, $\mu=Q$ and $\mu =2Q$. In each figure we see
that if $\mu$ gets larger the cross sections increase at
small $Q$ but decrease it at larger $Q$. Therefore there are specific
values in $Q$ where the scale variations are small and as we go from
LO to NLO to NNLO these points are at larger values of $Q$. 
The scale variation gets smaller as we go from
LO to NLO as expected. In spite of the fact that we do not have the
NNLO parton densities there is still an improvement in the scale
dependence if we go from NLO to NNLO.
The corresponding plots for the unpolarized cross
section are shown in Figs. 9a, 9b and 9c. These plots also show that when
$\mu$ becomes large the cross sections increase at small $Q$
and decrease it at larger $Q$. Further we see the expected overall
decrease in scale variation as we go from LO to NLO and then to NNLO.
  
To show the scale variation from a different point of view we plot the 
following
quantity
\begin{eqnarray}
\label{eq4.4}
N\left (\frac{\mu}{\mu_0}\right )=\frac{\Delta \sigma(\mu)}{\Delta 
\sigma(\mu_0)}\,.
\end{eqnarray}
In Fig. 10a we plot the polarized quantity in Eq. (\ref{eq4.4}) 
for $0.4<\mu/\mu_0<2$
at $\mu_0=Q=$ 5 GeV. In this figure we see  
an improvement in the scale variation while going from LO to NLO
and then to NNLO. 
This feature also persists at higher $Q$-values i.e. $Q=$ 10 GeV in Fig. 10b 
and $Q=$ 15 GeV in Fig. 10c.
For the unpolarized cross section the trend is the same. Here
we have the NNLO parton densities and
in Fig. 11a at $Q=$ 5 GeV we see an improvement in scale variation
while going to higher order. The same observation can be made in Fig. 11b at 
$Q=$ 10 GeV. However for $Q=$ 15 GeV in Fig. 11c the NLO curve becomes 
slightly worse than the LO one. While on the other hand the NNLO curve
is slightly better than the LO at least for larger scales $\mu$.

Now we look at effect of the higher order corrections on the longitudinal
asymmetry defined by
\begin{eqnarray}
\label{eq4.5}
A_{LL}=\frac{\Delta \sigma}{\sigma}\,.
\end{eqnarray}
In Fig. 12 we have plotted $A_{LL}$ in percent 
for BB1 in the range $2<Q<30$ GeV
in LO, NLO and NNLO. Since the polarized cross section changes sign 
we see that the asymmetry is negative for large $Q$ (note the displaced zero).
It is obvious that the longitudinal
asymmetry is small reaching about one percent between $Q=$ 10 GeV 
and $Q=$ 20 GeV.

To show the predictions of the various polarized parton density sets 
we compare the BB1 NLO contribution with the other three NLO 
parton density sets i.e. BB2, GRSV01 (standard scenario or SS) and GRSV01 
(valence scenario or VS) in Fig 13. Obviously the first three sets have
small positive longitudinal asymmetries while the GRSV01 valence scenario
set has a large negative asymmetry. Therefore the set with the smallest
gluon, here GRSV01 (valence scenario), has the largest asymmetry. On the other 
hand the set with the largest gluon, here BB1, has the smallest asymmetry.
This feature is unchanged in NNLO as can be seen in Fig. 14.
Hopefully the RHIC experiments will get sufficient luminosities and 
large enough proton polarizations to distinguish between the first 
three predictions and the last one.

To summarize we have computed the NNLO corrections to the production
of massive lepton-pairs in polarized proton-proton collisions.
Hence as far as coefficient functions as concerned we have the NNLO
corrections to both polarized and unpolarized processes. Unfortunately
the corresponding anomalous dimensions of the partons are not available
at NNLO. While fits have been made in the unpolarized case it is not possible
yet in the polarized case. Therefore we have convoluted the NLO polarized
densities with the NNLO coefficient functions to get an idea of the 
stability of the perturbation series. First we have seen that the 
dominance of the sea-quark initiated process, which holds at LO and NLO,
also prevails at NNLO. For the unpolarized cross section we have shown
that the addition of the NNLO terms produces a significant reduction in the
scale variation of the cross section. However the same phenomenon is
also shown for the polarized cross section in spite of the fact
that the NNLO parton densities are not known yet. 
Finally
the longitudinal asymmetry is small and positive for the polarized
parton density sets BB1, BB2 and GRSV(SS). If one uses the GRSV(VS)
densities then the longitudinal polarization is negative and large.


\appendix
\mysection*{Appendix A}
\setcounter{section}{1}
The lowest order contribution originating from the Born reaction Eq. 
(\ref{eq3.1}) is given by
\begin{eqnarray}
\label{eqA.1}
\Delta_{q\bar q}^{(0)}=-\delta(1-z)\,.
\end{eqnarray}
The O($\alpha_s$) correction to the $q\bar q$ subprocess in Eq. (\ref{eq3.2})
has been calculated
in the literature \cite{weber}-\cite{kamal}. 
Choosing the ${\overline {\rm MS}}$ scheme the 
expression for $\Delta_{q\bar q}^{(1)}$ can be obtained by using $n$-dimensional
regularization. For convenience we split the contribution as follows
\begin{eqnarray}
\label{eqA.2}
\Delta_{q\bar q}^{(1)}(z)=\Delta_{q\bar q}^{(1),{\rm S+V}}(z)+
\Delta_{q\bar q}^{(1),{\rm H}}(z)\,.
\end{eqnarray}
Here S+V indicates that the corrections are coming from soft-plus-virtual
gluons where H means that the corrections are coming from the hard $z\not =1$
gluon region. The expressions are equal to
\begin{eqnarray}
\label{eqA.3}
\Delta_{q\bar q}^{(1),{\rm S+V}}&=&-a_s\,C_F\,\Bigg \{
\delta(1-z)\Big [6\,\ln\left (\frac{Q^2}{\mu^2}\right )+8\,\zeta(2)
-16\Big ]
\nonumber\\[2ex]
&&+8\, {\cal D}_0(z)\,\ln\left (\frac{Q^2}{\mu^2}\right )+16\,{\cal D}_1
(z)\Bigg \}\,,
\end{eqnarray}
\begin{eqnarray}
\label{eqA.4}
\Delta_{q\bar q}^{(1),{\rm H}}&=&a_s\,C_F\,\Bigg \{4(1+z)\,\ln \left (
\frac{Q^2}{\mu^2}\right )+8(1+z)\,\ln(1-z)
\nonumber\\[2ex]
&&+4\frac{1+z^2}{1-z}\,\ln z\Bigg \}
\end{eqnarray}
where the distibutions ${\cal D}_i(z)$ are defined in Eq. (\ref{eq3.22}).
The second order correction to the non-singlet part $\Delta_{q\bar q}^{(2)}$
is determined by process in Eq. (\ref{eq3.4}) and the higher order corrections
to processes in Eq. (\ref{eq3.1}) and Eq. (\ref{eq3.2}). It can be split
into two parts. The first piece is related through mass factorization to
the collinearly singular part of the partonic structure function $W_{q\bar q}$
and will be denoted by
\begin{eqnarray}
\label{eqA.5}
\Delta_{\bar q}^{(2),{\rm NS}}=\Delta_{q\bar q}^{(2),{\rm S+V}}
+\Delta_{q\bar q}^{(2),{\rm C_A}}+\Delta_{q\bar q}^{(2),{\rm C_F}}+
\Delta_{q\bar q,A\bar A}^{(2)}+\Delta_{q\bar q,A\bar C}^{(2)}
+\Delta_{q\bar q,A\bar D}^{(2)}\,.
\end{eqnarray}
The second piece consists of the DY correction terms 
$\Delta_{q\bar q,B\bar B}^{(2)}$, $\Delta_{q\bar q,B\bar C}^{(2)}$,
$\Delta_{q\bar q,B \bar D}^{(2)}$. The contribution 
$\Delta_{q\bar q,A\bar B}^{(2)}$ is zero because of Furry's theorem.
First we will enumerate the contributions in Eq. (\ref{eqA.5}). The 
soft-plus-virtual gluon contributions come from the two-loop virtual graphs
contributing to the Born reaction in Eq. (\ref{eq3.1}), the soft gluon 
corrections to the reactions in Eqs. (\ref{eq3.2}), (\ref{eq3.4}) and soft 
quark pair production due to diagrams A in Fig. \ref{fig1}. The expression for 
this part is equal to
\begin{eqnarray}
\label{eqA.6}
\Delta_{q \bar q}^{(2),S+V}&=&-
a_s^2\, \delta(1-z)\, \Bigg\{
C_A\,C_F\,\Bigg[-11 \ln^2\left({Q^2 \over \mu^2}\right)
+\Big[{193 \over 3} 
\nonumber \\[2ex]
&&-24~\zeta(3)\Big ] \ln\left({Q^2 \over \mu^2}\right )
-{12 \over 5}~ \zeta(2)^2 +{592 \over 9}~ \zeta(2) +28 \zeta(3) 
- { 1535 \over 12} \Bigg ]
\nonumber \\[2ex]
&&
+C_F^2\,\Bigg[ \Big[18-32~\zeta(2)\Big ] 
\ln^2\left({Q^2 \over \mu^2}\right)
+\Big [24 \zeta(2)+176 \zeta(3) 
\nonumber \\[2ex]
&&
-93\Big]
\ln\left({Q^2 \over \mu^2}\right)
+{8 \over 5}~ \zeta(2)^2-70~\zeta(2)
\nonumber \\[2ex]
&&
-60~\zeta(3)+{511 \over 4}\Bigg ]
+n_f\,C_F\,\Bigg[2 \ln^2\left({Q^2 \over \mu^2}\right)-{34 \over 3} 
\ln\left({Q^2 \over \mu^2}\right)
\nonumber \\[2ex]
&&
+8~ \zeta(3)-{112 \over 9}~ \zeta(2)
+{127 \over 6}\Bigg ]\Bigg\}
+C_A\,C_F\,\Bigg[-{44 \over 3}~ {\cal D}_0(z) \ln^2\left({Q^2 \over \mu^2}
\right) \nonumber \\[2ex]
&&
+\Bigg [ \Bigg[{536 \over 9}-16~\zeta(2)\Bigg ]~ {\cal D}_0(z)
-{176 \over 3}~ {\cal D}_1(z)\Bigg ] 
\ln\left({Q^2 \over \mu^2}\right)
-{176 \over 3}~ {\cal D}_2(z) 
\nonumber \\[2ex]
&&
+ \Bigg[{1072 \over 9}-32~ \zeta(2)\Bigg ]~ {\cal D}_1(z)
+\Bigg[56 \zeta(3)+{176 \over 3}~ \zeta(2) -{1616 \over 27} \Bigg ]~ 
\nonumber \\[2ex]
&&
\times {\cal D}_0(z)\Bigg ]
+C_F^2\, \Bigg[\Big[64~ {\cal D}_1(z)+48~ {\cal D}_0(z)\Big ] 
\ln^2\left({Q^2 \over \mu^2}\right) 
\nonumber \\[2ex]
&&
+\Big[192~ {\cal D}_2(z)+96~ {\cal D}_1(z)-(128+64~ \zeta(2))~ 
{\cal D}_0(z)\Big ]
\ln\left({Q^2 \over \mu^2}\right)
\nonumber \\[2ex]
&&
+128~ {\cal D}_3(z)-(128~ \zeta(2)+256)~ {\cal D}_1(z)
+256 \zeta(3)~ {\cal D}_0(z)\Bigg ]
\nonumber \\[2ex]
&&
+n_f\,C_F\,\Bigg[{8 \over 3}~ {\cal D}_0(z) \ln^2\left({Q^2 \over \mu^2}\right)
+\Bigg[{32 \over 3}~ {\cal D}_1(z)-{80 \over 9}~ {\cal D}_0(z)\Bigg ] 
\nonumber \\[2ex]
&&
\times \ln\left({Q^2 \over \mu^2}\right)
+{32 \over 3}~ {\cal D}_2(z)-{160 \over 9}~ {\cal D}_1(z)
\nonumber \\[2ex]
&&
+({224 \over27}
-{32 \over 3}~ \zeta(2))~ {\cal D}_0(z)\Bigg ]\,.
\end{eqnarray}
The hard gluon contribitions are denoted by $\Delta_{q\bar q}^{(2),{\rm C_A}}$
and $\Delta_{q\bar q}^{(2),{\rm C_F}}$. They are equal to
\begin{eqnarray}
\label{eqA.7}
\Delta_{q \bar q}^{(2),C_A}\!\!&=&\!\!-
a_s^2\,C_A\,C_F\, \Bigg\{
{22 \over 3} (1+z) \ln^2\left({Q^2 \over \mu^2}\right)
+\Bigg[{1+z^2 \over 1-z}\Bigg[-8 {\rm Li}_2(1-z)
\nonumber \\[2ex]
&&
+{70 \over 3} \ln(z)\Bigg ]
+(1+z) \Bigg[8 \zeta(2) + {88 \over 3} \ln(1-z)
-6 \ln(z) \Bigg ]
-{4 \over 9}\Big(19
\nonumber \\[2ex]
&&
+124 z\Big) \Bigg ] \ln\left({Q^2 \over \mu^2}\right)
+{1+z^2 \over 1-z} \Bigg[-4 { \rm S}_{1,2}(1-z)-12 {\rm Li}_3(1-z)
\nonumber \\[2ex]
&&
+{4 \over 3} {\rm Li}_2(1-z)
+8 {\rm Li}_2(1-z) \ln(z)-8 {\rm Li}_2(1-z)\ln(1-z)
\nonumber \\[2ex]
&&
+8 \zeta(2) \ln(z) -{29 \over 2} \ln^2(z)
-{104 \over3} \ln(z)+{140 \over 3} \ln(z) \ln(1-z)\Bigg ]
\nonumber \\[2ex]
&&
+(1+z) \Bigg[16 { \rm S}_{1,2}(1-z)-12 {\rm Li}_3(1-z)-28 \zeta(3)
\nonumber \\[2ex]
&&
+8 {\rm Li}_2(1-z)\ln(1-z)
+16 \zeta(2)\ln(1-z)+{88 \over 3} \ln^2(1-z)
\nonumber \\[2ex]
&&
+{23\over6} \ln^2(z) - 12 \ln(z)\ln(1-z) \Bigg ]
\nonumber \\[2ex]
&&
-{4 \over 3 } (7+13 z) {\rm Li}_2(1-z)
-{4 \over 3} (19+25 z) \zeta(2) 
-{2 \over 3} (26-57 z) \ln(z)
\nonumber \\[2ex]
&&
-{4 \over 9} (38 +239 z ) \ln(1-z) 
-{446 \over 27} +{2278 \over 27} z\Bigg\}\,,
\end{eqnarray}
and 
\begin{eqnarray}
\label{eqA.8}
\Delta_{q \bar q}^{(2),C_F}\!\!&=&\!\!-
a_s^2\,C_F^2\,\Bigg \{
\Bigg[-16 {1 +z^2 \over 1-z} \ln(z)+8 (1+z)\Big[\ln(z)-4\ln(1-z)\Big ]
\nonumber \\[2ex]
&&
-8 (5+z)\Bigg ] \ln^2\left({Q^2 \over \mu^2}\right)+\Bigg[{1+z^2 \over 1-z}
\Big[ 16 {\rm Li}_2(1-z)+24 \ln^2(z)
\nonumber \\[2ex]
&&
-24 \ln(z) -112 \ln(z) \ln(1-z)\Big ]
+(1+z) \Big[32 {\rm Li}_2(1-z)
\nonumber \\[2ex]
&&
+32 \zeta(2)
-12 \ln^2(z)
+32\ln(z) \ln(1-z)-96 \ln^2(1-z)\Big ]
\nonumber \\[2ex]
&&
+8 (15+2 z) 
+16 (2-3 z) \ln(z)
-16 (7-z)\ln(1-z)\Bigg ]
\nonumber \\[2ex]
&&
\times \ln\left({Q^2 \over \mu^2}\right)
+{1+z^2 \over 1-z}\Bigg[-32 { \rm S}_{1,2}(1-z)-8 {\rm Li}_3(1-z)
\nonumber \\[2ex]
&&
-24 {\rm Li}_2(1-z) \ln(z)
+24 {\rm Li}_2(1-z) \ln(1-z)-12 \ln^3(z)
\nonumber \\[2ex]
&&
+64 \zeta(2) \ln(z)+72 \ln^2(z) \ln(1-z)-124 \ln^2(1-z) \ln(z)
\nonumber \\[2ex]
&&
+56 \ln(z)\Bigg ]
+(1-z)\Big[64 \zeta(2) -64\ln^2(1-z)\Big ]
\nonumber \\[2ex]
&&
+(1+z) \Bigg[{14 \over 3} \ln^3(z)-64 \ln^3(1-z)
-40 {\rm Li}_3(1-z)
\nonumber \\[2ex]
&&
+48 {\rm Li}_2(1-z) \ln(1-z)-32 \zeta(2) \ln(z)+16 { \rm S}_{1,2}(1-z)
\nonumber \\[2ex]
&&
+64 \zeta(2) \ln(1-z)-128 \zeta(3)-24 \ln^2(z) \ln(1-z) 
\nonumber \\[2ex]
&&
+32 \ln^2(1-z) \ln(z)\Bigg ]
-8 (3+4 z) {\rm Li}_2(1-z) 
\nonumber \\[2ex]
&&
-8 (1-3 z) \ln^2(z)
+16(4-7 z) \ln(z) \ln(1-z)
\nonumber \\[2ex]
&&
-8 (6-13 z) \ln(z) +4 (64+3 z) \ln(1-z)
-24 (3-2 z)\Bigg\}\,.
\end{eqnarray}
The function ${\rm Li}_n(z)$ and ${\rm S}_{n,p}(z)$ denote the polylogarithms 
and can be found in \cite{lewin}. The hard part of quark pair production due 
to the diagrams A in Fig. \ref{fig1} is equal to
\begin{eqnarray}
\label{eqA.9}
\Delta_{q \bar q,A \overline A}^{(2)}&=&-
a_s^2\,n_f\,C_F\, \Bigg\{
-{4 \over 3} (1+z) \ln^2\left({Q^2 \over \mu^2}\right)
+\Bigg[-{16 \over 3} {1+z^2 \over 1-z}
\nonumber \\[2ex]
&&
\times \ln(z) -{16 \over 3} (1+z) \ln(1-z) -{8 \over 9} (1-11 z) \Bigg ] 
\ln\left({Q^2 \over \mu^2}\right) 
\nonumber \\[2ex]
&&
+{1+z^2 \over 1-z} \Bigg[ 4 \ln^2(z)-{4 \over 3} {\rm Li}_2(1-z) + {20 \over 3 } \ln(z)
\nonumber \\[2ex]
&&
-{32 \over 3} \ln(z) \ln(1-z)\Bigg ] + (1+z) \Bigg[{4 \over 3 } {\rm Li}_2(1-z)
\nonumber \\[2ex]
&&
+{16 \over 3 } \zeta(2)-{16 \over 3} \ln^2(1-z) +{2 \over 3} \ln^2(z)\Bigg ]
-{16 \over 9} \Big(1-11 z\Big)
\nonumber \\[2ex]
&&
\times \ln(1-z) +{8 \over 3}\Big(2-3 z\Big) \ln(z) +{4 \over 27}
\Big(47 -103 z\Big)
\Bigg\}\,,
\end{eqnarray}
Finally, we have the interference terms corresponding to the combinations
$A\bar C$ and $A\bar D$ in Fig. \ref{fig1} and Fig. \ref{fig2}. For them
we find
\begin{eqnarray}
\label{eqA.10}
\Delta_{q \bar q,A \overline C}^{(2)}&=&
\Delta_{q \bar q,A \overline D}^{(2)}=-
a_s^2\, C_F\,\left(C_F-{1 \over 2}\,C_A\right)\, \Bigg\{
\Bigg[{1+z^2 \over 1-z}\Bigg[-8 {\rm Li}_2(1-z)
\nonumber \\[2ex]
&&
-4 \ln^2(z)-6 \ln(z)\Bigg ]-14 (1+z) \ln(z)-4(8-7 z) \Bigg ] 
\ln\left({Q^2 \over \mu^2}\right) 
\nonumber \\[2ex]
&&
+{1+z^2 \over 1-z} \Bigg[16 {\rm Li}_3(1-z)-36 { \rm S}_{1,2}(1-z)
+{8 \over 3}\ln^3(z)
\nonumber \\[2ex]
&&
-12 {\rm Li}_2(1-z) \ln(z)-16 {\rm Li}_2(1-z) \ln(1-z)
\nonumber \\[2ex]
&&
-6 {\rm Li}_2(1-z)
+{15 \over 2 } \ln^2(z)-8 \ln^2(z) \ln(1-z)
\nonumber \\[2ex]
&&
+12 \ln(z) -12 \ln(z) \ln(1-z)\Bigg ]
+(1+z) \Bigg[-8 {\rm Li}_3(1-z)
\nonumber \\[2ex]
&&
-26 {\rm Li}_2(1-z)+4 {\rm Li}_2(1-z) \ln(z) 
+{2 \over 3} \ln^3(z)
+{23 \over 2} \ln^2(z)
\nonumber \\[2ex]
&&
-28 \ln(z) \ln(1-z)\Bigg ]
+2 (22-9 z) \ln(z)-8 (8-7 z) \ln(1-z)
\nonumber \\[2ex]
&&
+2 (47 -39 z)\}\,.
\end{eqnarray}
The remaining parts of $q\bar q$ scattering are free of mass singularities.
Therefore they do not need mass factorization, which implies that their
contributions are scheme and scale independent. The contributions from the 
diagrams B in Fig. \ref{fig1} and the interference terms $B\bar C$ and 
$B\bar D$ (see Figs. \ref{fig1}, \ref{fig2}) are
\begin{eqnarray}
\label{eqA.11}
\Delta_{q \bar q,B \overline B}^{(2)}&=&-
a_s^2\,T_f\,C_F\,\Bigg\{
(1+z)^2 \Bigg[-{32 \over 3 } {\rm Li}_2(-z)-{16 \over 3} \zeta(2)
+{8 \over 3} \ln^2(z)
\nonumber \\[2ex]
&&
-{32 \over 3 } \ln(z) \ln(1+z)\Bigg ]+{8 \over 3} \Big(3+3 z^2+4 z\Big) \ln(z)
\nonumber \\[2ex]
&&
+{40 \over 3}\Big(1-z^2\Big)\Bigg \}\,,
\end{eqnarray}
and
\begin{eqnarray}
\label{eqA.12}
\Delta_{q \bar q,B \overline C}^{(2)}&=&
\Delta_{q \bar q,B \overline D}^{(2)}=-
a_s^2\, C_F\,\left(C_F-{1 \over 2}\, 
C_A\right)\,\Bigg\{(1+z^2+3 z)
\nonumber \\[2ex]
&&
\times \Bigg[ 32 { \rm S}_{1,2}(1-z)
+16 {\rm Li}_2(1-z) \ln(z)\Bigg ]
+(1+z)^2 \Bigg[-48 { \rm S}_{1,2}(-z)
\nonumber \\[2ex]
&&
-8 {\rm Li}_3(-z)+24 {\rm Li}_2(-z)+24 {\rm Li}_2(-z) \ln(z)
\nonumber \\[2ex]
&&
-48 {\rm Li}_2(-z) \ln(1+z)+12 \zeta(2)-24 \zeta(2) \ln(1+z)
\nonumber \\[2ex]
&&
+8 \zeta(2) \ln(z)+20 \ln^2(z) \ln(1+z)-24 \ln^2(1+z) 
\nonumber \\[2ex]
&&
\times \ln(z)+24 \ln(z) \ln(1+z)\Bigg ]
+36 (1-z^2) {\rm Li}_2(1-z)
\nonumber \\[2ex]
&&
+{4 \over 3} \Big(1+z^2+4 z\Big) \ln^3(z) +4 (9+11 z) \ln(z)
-2 (-6
\nonumber \\[2ex]
&&
+15 z^2+8 z) \ln^2(z)-2 (-27
+13 z^2+14 z) \Bigg \}\,.
\end{eqnarray}
Because of Furry's theorem
\begin{eqnarray}
\label{eqA.13}
\Delta_{q\bar q,A\overline B}^{(2)}=0\,.
\end{eqnarray}
At O($\alpha_s$) the $qg$ subprocess shows up for the first time. The Drell-Yan
correction term for the reaction in Eq. (\ref{eq3.3}) has been calculated in 
\cite{weber}, \cite{mara}, \cite{gehr}, \cite{kamal} and it is given by
\begin{eqnarray}
\label{eqA.14}
\Delta_{qg}^{(1)}=-a_s\,T_f\,\Bigg \{\Big [4\,z-2\Big ]\,\ln\left (\frac{
(1-z)^2\,Q^2}{z\,\mu^2}\right )+(1-z)(5+3\,z)\Bigg \}\,.
\end{eqnarray}
The second order part of $\Delta_{qg}$ receives contributions from the
virtual corrections to the reaction in Eq. (\ref{eq3.3}) and the process 
in Eq. (\ref{eq3.5})
\begin{eqnarray}
\label{eqA.15}
\Delta_{qg}^{(2)}=\Delta_{\bar qg}^{(2)}=\Delta_{qg}^{(2),{\rm C_A}}
+\Delta_{qg}^{(2),{\rm C_F}}\,.
\end{eqnarray}
The calculation of $\Delta_{qg}^{(2)}$ requires both mass factorization
and renormalization. The two parts $\Delta_{qg}^{(2),{\rm C_A}}$
and $\Delta_{qg}^{(2),{\rm C_F}}$ are equal to
\begin{eqnarray}
\label{eqA.16}
\Delta_{q g}^{(2),C_A}&=&
a_s^2\,C_A\,T_f\,\Bigg\{
\Big[ -8 (1+z ) \ln(z) +4 (1-2 z) \ln(1-z)
\nonumber \\[2ex]
&&
-24 (1-z) \Big]
\ln^2\left({Q^2 \over \mu^2}\right)
+\Big[ 8 (1+ 2 z)  \Big[ {\rm Li}_2(-z) 
\nonumber \\[2ex]
&&
+ \ln(z) \ln(1+z)-3 {\rm Li}_2(1-z)\Big]
+4 (3 +4 z) \ln^2(z)
\nonumber \\[2ex]
&&
-8 (1-4 z) \zeta(2)
-8 (5 +2 z) \ln(z) \ln(1-z)
+12 (1-2 z) 
\nonumber \\[2ex]
&&
\times \ln^2(1-z)
-4 (25-22 z-3 z^2) \ln(1-z)
\nonumber \\[2ex]
&&
+4 (13 -22 z -3 z^2) \ln(z)
\nonumber \\[2ex]
&&
+4 (17-18 z) \Big]
\ln\left({Q^2 \over \mu^2}\right)
-24 (1+z) \Big[{\rm Li}_2(-z)
\nonumber \\[2ex]
&&
+\ln(z) \ln(1+z)\Big]
+4 (1+2 z) \Big[ 2 {\rm Li}_3(-z) 
\nonumber \\[2ex]
&&
+ 4 {\rm Li}_3\left(-{1 -z \over 1+z} \right)
-4 {\rm Li}_3\left( {1-z \over 1+z}\right)
-4 \ln(z) {\rm Li}_2(-z)
\nonumber \\[2ex]
&&
+4 \ln(1-z) {\rm Li}_2(-z)
-3 \ln^2(z) \ln(1+z)
+4 \ln(z) 
\nonumber \\[2ex]
&&
\times \ln(1-z) \ln(1+z) \Big]
-8 (11 +12 z) { \rm S}_{1,2}(1-z)
\nonumber \\[2ex]
&&
+4 (19+26 z) {\rm Li}_3(1-z)
+8 (1+z) \zeta(3)
-8 (4-z) \ln(z) 
\nonumber \\[2ex]
&&
\times {\rm Li}_2(1-z)
-48 (1+2 z) \ln(1-z) {\rm Li}_2(1-z)
\nonumber \\[2ex]
&&
+4 (-29 +8 z -6 z^2) {\rm Li}_2(1-z)
+40 \zeta(2) \ln(z)
\nonumber \\[2ex]
&&
-16 (1-4 z) \zeta(2) \ln(1-z)
+4 (29 -38 z +3 z^2) \zeta(2)
\nonumber \\[2ex]
&&
-{2 \over 3} (11 +16 z) \ln^3(z)
+{26 \over 3} (1-2 z) \ln^3(1-z)
\nonumber \\[2ex]
&&
+(-47+92 z +12 z^2) \ln^2(z)
+4 (7+6 z) \ln^2(z) \ln(1-z)
\nonumber \\[2ex]
&&
-4 (11+2 z) \ln(z) \ln^2(1-z)
+4(-26 +23 z+3 z^2) 
\nonumber \\[2ex]
&&
\times \ln^2(1-z)
+4 (-36+35 z) \ln(z)
+4 (27-44 z -9 z^2)
\nonumber \\[2ex]
&&
\times \ln(z) \ln(1-z)
+2(-72 +71 z+3 z^2)
\nonumber \\[2ex]
&&
+2 (68-71 z) \ln(1-z)\Bigg \}\,,
\end{eqnarray}
and
\begin{eqnarray}
\label{eqA.17}
\Delta_{qg}^{(2),C_F}&=&
a_s^2\,C_F\,T_f\, \Bigg\{
\Big[12 (1-2 z ) \ln(1-z)
-6 (1-2 z) \ln(z)
\nonumber \\[2ex]
&&
-9 \Big] \ln^2\left({Q^2 \over \mu^2}\right)
+\Big[4 (1-2 z) \Big[9 \ln^2(1-z) 
-2 \zeta(2) +2 \ln^2(z)
\nonumber \\[2ex]
&&
-10 \ln(z) \ln(1-z) \Big]
+2 (29-8 z -6 z^2) \ln(z)
\nonumber \\[2ex]
&&
+6 (-16 +12 z+2 z^2) \ln(1-z)
+2 (49-38 z) \Big]
\nonumber \\[2ex]
&&
\times \ln\left({Q^2 \over \mu^2}\right)
+32 (1+z) \Big[{\rm Li}_2(-z)+\ln(z) \ln(1+z)\Big]
\nonumber \\[2ex]
&&
+2 (1-2 z) 
\Big[16 {\rm Li}_3(-z)
+50 \zeta(3)
\nonumber \\[2ex]
&&
-8 {\rm Li}_2(-z) \ln(z)
-8 \zeta(2) \ln(1-z)
\nonumber \\[2ex]
&&
+{35 \over 3} \ln^3(1-z) \Big]
-12 (1-2 z) { \rm S}_{1,2}(1-z)
-20 (1-2 z) 
\nonumber \\[2ex]
&&
\times {\rm Li}_3(1-z)
+4 (1-2 z) \ln(z) {\rm Li}_2(1-z)
+4 (1-2 z)
\nonumber \\[2ex]
&&
\times  \ln(1-z) 
{\rm Li}_2(1-z)
+ 2 (13+28 z) {\rm Li}_2(1-z)
\nonumber \\[2ex]
&&
+24 (1-2 z) \zeta(2) \ln(z)
+4 (1+10 z-6 z^2) \zeta(2) 
\nonumber \\[2ex]
&&
-{17 \over 3} (1-2 z) \ln^3(z)
+24 (1-2 z) \ln(1-z) \ln^2(z)
\nonumber \\[2ex]
&&
+(-{81\over 2} -20 z+6 z^2) \ln^2(z)
-42 (1-2 z) \ln^2(1-z) \ln(z)
\nonumber \\[2ex]
&&
+2 (-55+52 z+9 z^2) \ln^2(1-z)
+4 (35-12 z-6 z^2) 
\nonumber \\[2ex]
&&
\times \ln(z) \ln(1-z)
+(-175 + 79 z +60 z^2) \ln(z)
\nonumber \\[2ex]
&&
-2(-128+77 z 
+30 z^2) \ln(1-z)
-{627 \over 2} +331 z-{15 \over 2} z^2\Bigg \}\,.
\nonumber\\[2ex]
\end{eqnarray}
The reactions in Fig. \ref{fig2} describe quark-anti-quark as well as 
quark-quark scattering (without identical quarks). The contribution to the
DY correction term can be split into two parts. The first part, represented
by the combinations $C\bar C$ and $D\bar D$ needs mass factorization.
In this case the contributions for $q\bar q$, $qq$ and $\bar q\bar q$
are all equal and are given by
\begin{eqnarray}
\label{eqA.18}
\Delta_{q \bar q,C \overline C}^{(2)}&=&
\Delta_{q \bar q,D \overline D}^{(2)}=
\Delta_{q q,C \overline C}^{(2)}=
\Delta_{q q,D \overline D}^{(2)}=
\Delta_{\bar q \bar q,C \overline C}^{(2)}=
\Delta_{\bar q \bar q,D \overline D}^{(2)}=
\nonumber \\[2ex]
&&
a_s^2\,C_F\,T_f\,\Bigg\{
\Big[-4 (1+z) \ln(z) -10 (1-z) \Big] 
\ln^2\left({Q^2 \over \mu^2}\right)
\nonumber \\[2ex]
&&
+\Big[8 (1+z) \Big[-2 {\rm Li}_2(1-z) + \ln^2(z) 
-2 \ln(z) \ln(1-z) \Big]
\nonumber \\[2ex]
&&
+4 (7 - 8 z) \ln(z)
-40 (1-z) \ln(1-z)
+40 (1-z)\Big]
\nonumber \\[2ex]
&&
\times \ln\left({Q^2 \over \mu^2}\right)
-40 (1-z) \Big[\ln^2(1-z) -\zeta(2)\Big]
+8 (1+z) 
\nonumber \\[2ex]
&&
\times \Big[-6 { \rm S}_{1,2}(1-z)
+4 {\rm Li}_3(1-z)
-\ln(z) {\rm Li}_2(1-z)
\nonumber \\[2ex]
&&
-4 \ln(1-z) {\rm Li}_2(1-z)
+2 \zeta(2) \ln(z)
-{3 \over 4} \ln^3(z)
\nonumber \\[2ex]
&&
+2 \ln^2(z) \ln(1-z)
-2 \ln(z) \ln^2(1-z) \Big]
-4 (9-7 z) 
\nonumber \\[2ex]
&&
\times {\rm Li}_2(1-z)
+8 (7-8 z) \ln(z) \ln(1-z)
+(-25 +39 z) 
\nonumber \\[2ex]
&&
\times \ln^2(z)
+80 (1-z) \ln(1-z)
+2 (-42 +17 z) \ln(z)
\nonumber \\[2ex]
&&
-131 +128 z +3 z^2\Bigg \}\,.
\end{eqnarray}
The second part, which is collinearly finite, consists of the interference
between the graphs C and D in Fig. \ref{fig2}. Further notice the relative
minus sign between the $q\bar q$ and the $qq$ ($\bar q\bar q$) part.
The expresions for these interference terms are
\begin{eqnarray}
\label{eqA.19}
\Delta_{q \bar q C \overline D}^{(2)}&=&
-\Delta_{q  q C \overline D}^{(2)}=
-\Delta_{\bar q \bar q C \overline D}^{(2)}=
\nonumber \\[2ex]
&&
a_s^2\,C_F\,T_f\, \Bigg \{
8 (2+z) \Big[-4 { \rm S}_{1,2}(1-z)
+12 { \rm S}_{1,2}(-z)
\nonumber \\[2ex]
&&
+6 \ln^2(1+z) \ln(z)
+6 \zeta(2) \ln(1+z)
-5 \ln(1+z) \ln^2(z)
\nonumber \\[2ex]
&&
+12 \ln(1+z) {\rm Li}_2(-z) \Big]
-8 (1+z) \Big[2 {\rm Li}_2(-z)
+2 \ln(z) \ln(1+z)
\nonumber \\[2ex]
&&
+ \zeta(2) \Big]
-8 (2-z ) {\rm Li}_3(1-z)
+16 (6- 5 z) {\rm Li}_3(-z)
\nonumber \\[2ex]
&&
+8(6-9 z) \zeta(3)
-8 (2+3 z) {\rm Li}_2(1-z) \ln(z)
+{16 \over 3} z \ln^3(z)
\nonumber \\[2ex]
&&
-128 \ln(z) {\rm Li}_2(-z)
-8(2+5 z) \zeta(2) \ln(z)
-8 {\rm Li}_2(1-z)
\nonumber \\[2ex]
&&
+4 z \ln^2(z)
+16 \ln(z)
+32 (1-z)\Bigg \}\,.
\end{eqnarray}
In case there are identical quarks in the final state, we have in addition
to the graphs in Fig. \ref{fig2} also the ones in Fig. \ref{fig3}. As the 
results for $E\bar E$, $F\bar F$ and $E\bar F$ are equal to those for
$C\bar C$, $D\bar D$ and $C\bar D$ (of course one has to implement the right
statistical factors), we will not discuss them here. The new contributions
come from the intereference terms $C\bar E$, $C\bar F$, $D\bar E$ and 
$D\bar F$. Further when the intermediate vector boson is a photon all four sets of diagrams C,D,E and F contribute and we have a statistical factor 1/2.
Apart from the statistical factor there is another difference bewtween
$C\bar E$ ($D\bar F$) and $C\bar F$ ($D \bar E$). The first contains
collinear divergences and needs mass factorization, whereas the latter is free 
of mass singularities. The correction corresponding to the interferences
$C\bar E$ and $D\bar F$ is equal to
\begin{eqnarray}
\label{eqA.20}
\Delta_{q  q,C \overline E}^{(2)}&=&\Delta_{q q,D \overline F}^{(2)}=
\Delta_{\bar q \bar q,C \overline E}^{(2)}=\Delta_{\bar q \bar q,
D \overline F}^{(2)}=
a_s^2  C_F\left(C_F-{1 \over 2} C_A\right)\Bigg \{
\Bigg[{1+z^2 \over 1+z} 
\nonumber \\[2ex]
&&
\times \Bigg[ 4 \ln^2(z)-8 \zeta(2)-16 {\rm Li}_2(-z)-16 \ln(z) \ln(1+z)\Bigg ]
\nonumber \\[2ex]
&&
+8 (1+z) \ln(z)+16 (1-z)\Bigg ]\ln\left({Q^2 \over \mu^2}\right)
+{1+z^2 \over 1+z} 
\nonumber \\[2ex]
&&
\times \Bigg[ 32 { \rm S}_{1,2}(1-z) -16 { \rm S}_{1,2}(-z)-32 {\rm Li}_3(1-z)
\nonumber \\[2ex]
&&
-8 {\rm Li}_3(-z)
-32 {\rm Li}_3\left(-{1-z \over 1+z}\right)+32 {\rm Li}_3
\left({1-z \over 1+z}\right)-4 \zeta(3)
\nonumber \\[2ex]
&&
+24 {\rm Li}_2(1-z) \ln(z)
+32 {\rm Li}_2(-z) \ln(z) -32 {\rm Li}_2(-z) \ln(1-z)
\nonumber \\[2ex]
&&
-16 {\rm Li}_2(-z) \ln(1+z)
+12 \zeta(2) \ln(z)-16 \zeta(2) \ln(1-z) 
\nonumber \\[2ex]
&&
-8 \zeta(2) \ln(1+z)-{8 \over 3} \ln^3(z)
+8 \ln^2(z)\ln(1-z)
\nonumber \\[2ex]
&&
+28 \ln^2(z) \ln(1+z)-8 \ln^2(1+z) \ln(z)
-32 \ln(z) \ln(1-z)
\nonumber \\[2ex]
&&
\times \ln(1+z)\Bigg ] +(1-z) \Bigg[-16 { \rm S}_{1,2}(-z) +8 {\rm Li}_3(-z)
+8 \zeta(3)
\nonumber \\[2ex]
&&
-16 {\rm Li}_2(-z) \ln(1+z) +4 \zeta(2) \ln(z)-8 \zeta(2) \ln(1+z)
\nonumber \\[2ex]
&&
-{2 \over 3} \ln^3(z)+4 \ln^2(z)\ln(1+z)-8 \ln^2(1+z) \ln(z)
\nonumber \\[2ex]
&&
+32\ln(1-z)-34\Bigg ]
+(1+z)\Bigg[8 {\rm Li}_2(-z)+4 \zeta(2)
\nonumber \\[2ex]
&&
+16 \ln(z) \ln(1-z)+8 \ln(z) \ln(1+z)\Bigg ]
+8 (3+z) 
\nonumber \\[2ex]
&&
\times {\rm Li}_2(1-z)
-4 (1+3 z) \ln^2(z)-2 (9-7 z) \ln(z)\Bigg\}\,.
\end{eqnarray}
The expression for the interference terms $C\bar F$ and $D\bar E$ is
\begin{eqnarray}
\label{eqA.21}
\Delta_{q  q,C \overline F}^{(2)}&=&\Delta_{q q,D \overline E}^{(2)}=
\Delta_{\bar q \bar q,C \overline F}^{(2)}=\Delta_{\bar q \bar q,D 
\overline E}^{(2)}=
\nonumber \\[2ex]
&&
a_s^2\,C_F(C_F-{1 \over 2} C_A)\, \Bigg\{
(1-z)^2\Bigg[8 {\rm Li}_3(1-z)
\nonumber \\[2ex]
&&
-12 {\rm Li}_2(1-z)-8 {\rm Li}_2(1-z) \ln(z)
-8 { \rm S}_{1,2}(1-z)
\nonumber \\[2ex]
&&
-{4 \over 3 } \ln^3(z)-6 \ln^2(z) \Bigg ] -2 (7 -6 z) \ln(z) 
\nonumber \\[2ex]
&&
- (15+13 z^2-28 z) \Bigg\}\,.
\end{eqnarray}
Notice that the above expression is scheme and scale independent.
The diagrams for the gluon-gluon subprocess can be obtained from the 
quark-anti-quark annihilation graphs via crossing. The subprocess shows up
for the first time at O($\alpha_s^2$). We have divided its Drell-Yan correction 
term into two parts i.e.
\begin{eqnarray}
\label{eqA.22}
\Delta_{gg}^{(2)}=\Delta_{gg}^{(2),{\rm C_A}}+\Delta_{gg}^{(2),{\rm C_F}}\,.
\end{eqnarray}
The $C_A$-contribution is collinearly finite and is therefore scheme and scale
independent. It is given by
\begin{eqnarray}
\label{eqA.23}
\Delta_{gg}^{(2),C_A}&=&
a_s^2\,{N^2 \over N^2-1}\, \Bigg\{
-(1+z)^2~\Big [ 16 { \rm S}_{1,2}(-z)+24 {\rm Li}_3(-z)
\nonumber \\[2ex]
&&
+16~\zeta(3)+{16 \over 3}~{\rm Li}_2(-z)
-24~{\rm Li}_2(-z) \ln(z)
\nonumber \\[2ex]
&&
+16 {\rm Li}_2(-z)\ln(1+z) + 8 \zeta(2) \ln(1+z)
+{8 \over 3} \zeta(2) 
\nonumber \\[2ex]
&&
-12 \ln^2(z) \ln(1+z) + 8 \ln^2(1+z) \ln(z)
+{16 \over 3} 
\ln(z) 
\nonumber \\[2ex]
&&
\times \ln(1+z) \Big]
-8 (1-z)^2 { \rm S}_{1,2}(1-z)+\Big(4 + {124 \over 3 } z + 22 z^2\Big) 
\nonumber \\[2ex]
&&
\times \ln(z)
+\Big({4 \over 3} -{28 \over 3} z -{14 \over 3} z^2\Big) \ln^2(z)
+{101 \over 3} (1-z^2) \Bigg\}\,.
\end{eqnarray}
The $C_F$-contribution contains collinear singularities. After mass 
factorization in the ${\overline {\rm MS}}$-scheme we find
\begin{eqnarray}
\label{eqA.24}
\Delta_{gg}^{(2),C_F}&=&
a_s^2\,  \Bigg\{
\Big[2 (1+4 z) \ln(z) + 8 (1-z) \Big] 
\ln^2\left({Q^2 \over \mu^2}\right)
\nonumber \\[2ex]
&&
+\Big[2 (1+4 z) (4 \ln(z) \ln(1-z) 
- \ln^2(z) + 4 {\rm Li}_2(1-z)) 
\nonumber \\[2ex]
&&
+32 (1-z) \ln(1-z) -2 (9-8 z) \ln(z)
-43+52 z - 9 z^2 \Big] 
\nonumber \\[2ex]
&&
\times \ln\left({Q^2 \over \mu^2}\right)
+4 (1+4 z) \Big[-4 {\rm Li}_3(1-z) + {\rm Li}_2(1-z) \ln(z) 
\nonumber \\[2ex]
&&
+4 {\rm Li}_2(1-z) \ln(1-z)
-\ln^2(z) \ln(1-z) 
+2 \ln(z) 
\nonumber \\[2ex]
&&
\times \ln^2(1-z) \Big]
-8 (1+z) \Big[{\rm Li}_2(-z)+\ln(z) \ln(1+z)\Big]
\nonumber \\[2ex]
&&
+4 (1+z)^2 \Big[4 { \rm S}_{1,2}(-z) + 4 {\rm Li}_2(-z) \ln(1+z)
\nonumber \\[2ex]
&&
+2 \zeta(2) \ln(1+z) -3 \ln^2(z) \ln(1+z) +2 \ln^2(1+z) 
\nonumber \\[2ex]
&&
\times\ln(z) \Big]
+8 (3 + 6 z +z^2) { \rm S}_{1,2}(1-z)
+8 (1+2 z -z^2) 
\nonumber \\[2ex]
&&
\times {\rm Li}_3(-z)
+4 (1+ 2 z -2 z^2) \zeta(3)
+4 (3-4 z) {\rm Li}_2(1-z)
\nonumber \\[2ex]
&&
-8 (2 +4 z + z^2) \ln(z) {\rm Li}_2(-z)
-4 (3 +10 z + 2 z^2) 
\nonumber \\[2ex]
&&
\times \zeta(2) \ln(z)
-4 (9- 7 z) \zeta(2)
+2 (1+ {8\over 3} z + {4 \over 3 }z^2) \ln^3(z)
\nonumber \\[2ex]
&&
+2 (7+3 z) \ln^2 z
+32 (1-z) \ln^2(1-z)
-4 (9-8 z)  
\nonumber \\[2ex]
&&
\times \ln(z) \ln(1-z)
+(55 - 72 z +27 z^2) \ln(z)
\nonumber \\[2ex]
&&
-2 (43-52 z +9 z^2) \ln(1-z)
+2 (35-31 z-4 z^2)
\Bigg\}\,.
\end{eqnarray}

%

\centerline{\bf \large{Figure Captions}}
%
\begin{description}
\item[Fig. 4.]
The NLO differential cross section $d\Delta \sigma/dQ$
at $\sqrt{s}=200$ GeV for the BB1 parton densities
plotted in the range $2<Q<30$ GeV
and $\mu^2=Q^2$. 
The NLO plots are for the O($\alpha_s$) corrected $q\bar q$ (dot-dashed line),
the O($\alpha_s$) contribution $qg$ (dashed line) and the total sum 
(solid line).  
Note that all contributions have negative regions at large $Q$ so we
have plotted their absolute values (see text).
\item[Fig. 5.]
Same as Fig. 4 but now for NNLO. Further we have shown the O($\alpha_s^2$)
corrected $q\bar q$ (dot-dashed line), the O($\alpha_s^2$) corrected $qg$ 
(long-dashed line), the O($\alpha_s^2$) contributions $gg$ (dotted line) and 
$qq$ (short-dashed line) and the total sum (solid line).
Note that all contributions have negative regions so we
have plotted their absolute values (see text).
\item[Fig. 6.]
The cross section $d\Delta \sigma/dQ$ at $\sqrt{s}=200$ GeV plotted in
range $2<Q<30$ GeV; LO (dotted line), NLO (dashed line), NNLO
(solid line).
Note that all contributions have negative regions at large $Q$ so we
have plotted their absolute values (see text).
\item[Fig. 7.]
The K-factor at $\sqrt{s}=200$ GeV plotted in the region $2<Q<30$ GeV;
$K^{\rm NLO}$ (dashed line), $K^{\rm NNLO}$ (solid line).
\item[Fig. 8a.]
The cross section $d\Delta \sigma/dQ$ at $\sqrt{s}=200$ GeV plotted in the 
region $2<Q<30$ GeV; LO $\mu=Q$ (solid line), LO $\mu=Q/2$ (dashed line)
, LO $\mu=2Q$ (dotted line). 
\item[Fig. 8b.]
The cross section $d\Delta \sigma/dQ$ at $\sqrt{s}=200$ GeV plotted in the
region $2<Q<30$ GeV. NLO $\mu=Q$ (solid line), NLO $\mu=Q/2$ 
(dashed line), NLO $\mu=2Q$ (dotted line). 
\item[Fig. 8c.]
The cross section $d\Delta \sigma/dQ$ at $\sqrt{s}=200$ GeV plotted in the
region $2<Q<30$ GeV. NNLO $\mu=Q$ (solid line), NNLO $\mu=Q/2$ 
(dashed line), NNLO $\mu=2Q$ (dotted line). 
\item[Fig. 9a.]
The unpolarized cross section $d\sigma/dQ$ at $\sqrt{s}=200$ GeV plotted in the
region $2<Q<30$ GeV. LO $\mu=Q$ (solid line), LO $\mu=Q/2$ 
(dashed line), LO $\mu=2Q$ (dotted line). 
\item[Fig. 9b.]
The unpolarized cross section $d\sigma/dQ$ at $\sqrt{s}=200$ GeV plotted in the
region $2<Q<30$ GeV. NLO $\mu=Q$ (solid line), NLO $\mu=Q/2$ 
(dashed line), NLO $\mu=2Q$ (dotted line).   
\item[Fig. 9c.]
The unpolarized cross section $d\sigma/dQ$ at $\sqrt{s}=200$ GeV plotted in the
region $2<Q<30$ GeV. NNLO $\mu=Q$ (solid line), NNLO $\mu=Q/2$ 
(dashed line), NNLO $\mu=2Q$ (dotted line).                                     
\item[Fig. 10a.]
The polarized quantity $N(\mu/\mu_0)$ (Eq. (\ref{eq4.4})) plotted in the range
$0.4<\mu/\mu_0<2$ with $Q=5$ GeV and $\mu_0=Q$. The results
are shown for LO (dotted line), NLO (dashed line), NNLO (solid line).
\item[Fig. 10b.]
Same as in Fig. 10a but now for $Q=10$ GeV.
\item[Fig. 10c.]
Same as in Fig. 10a but now for $Q=15$ GeV.
\item[Fig. 11a.]
The unpolarized quantity $N(\mu/\mu_0)$ (Eq. (\ref{eq4.4})) plotted in the 
range
$0.4<\mu/\mu_0<2$ with $Q=5$ GeV and $\mu_0=Q$. The results
are shown for LO (dotted line), NLO (dashed line), NNLO (solid line).
\item[Fig. 11b.]
Same as in Fig. 11a but now for $Q=10$ GeV.
\item[Fig. 11c.]
Same as in Fig. 11a but now for $Q=15$ GeV.
\item[Fig. 12.]
The longitudinal asymmetry $A_{\rm LL}$ (Eq. (\ref{eq4.5})) in $\%$ plotted
in the range $2<Q<30$ GeV at $\mu=Q$. LO (dotted line), NLO
(dashed line), NNLO (solid line).
\item[Fig. 13.]
The longitudinal asymmetry $A_{\rm LL}$ (Eq. (\ref{eq4.5})) in $\%$ plotted
in the range $2<Q<30$ GeV at $\mu=Q$ and NLO. BB1 (solid line),
BB2 (dot-dashed line), GRVS01, standard scenario (dashed line),
GRSV01, valence scenario (dotted line).
\item[Fig. 14.]
The longitudinal asymmetry $A_{\rm LL}$ (Eq. (\ref{eq4.5})) in $\%$ plotted
in the range $2<Q<30$ GeV at $\mu=Q$ and NNLO. BB1 (solid line),
BB2 (dot-dashed line), GRVS01, standard scenario (dashed line),
GRSV01, valence scenario (dotted line).
\end{description}

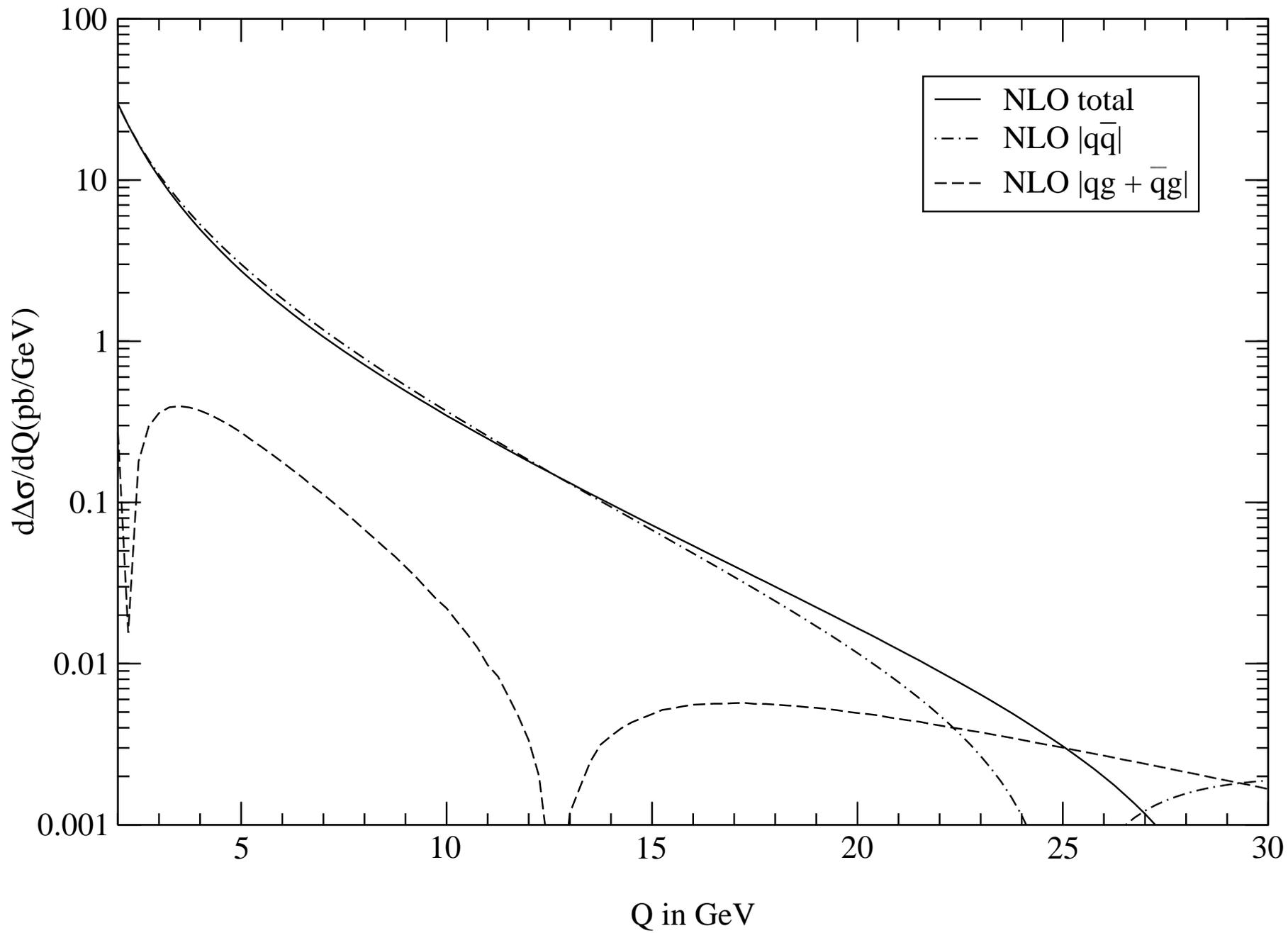
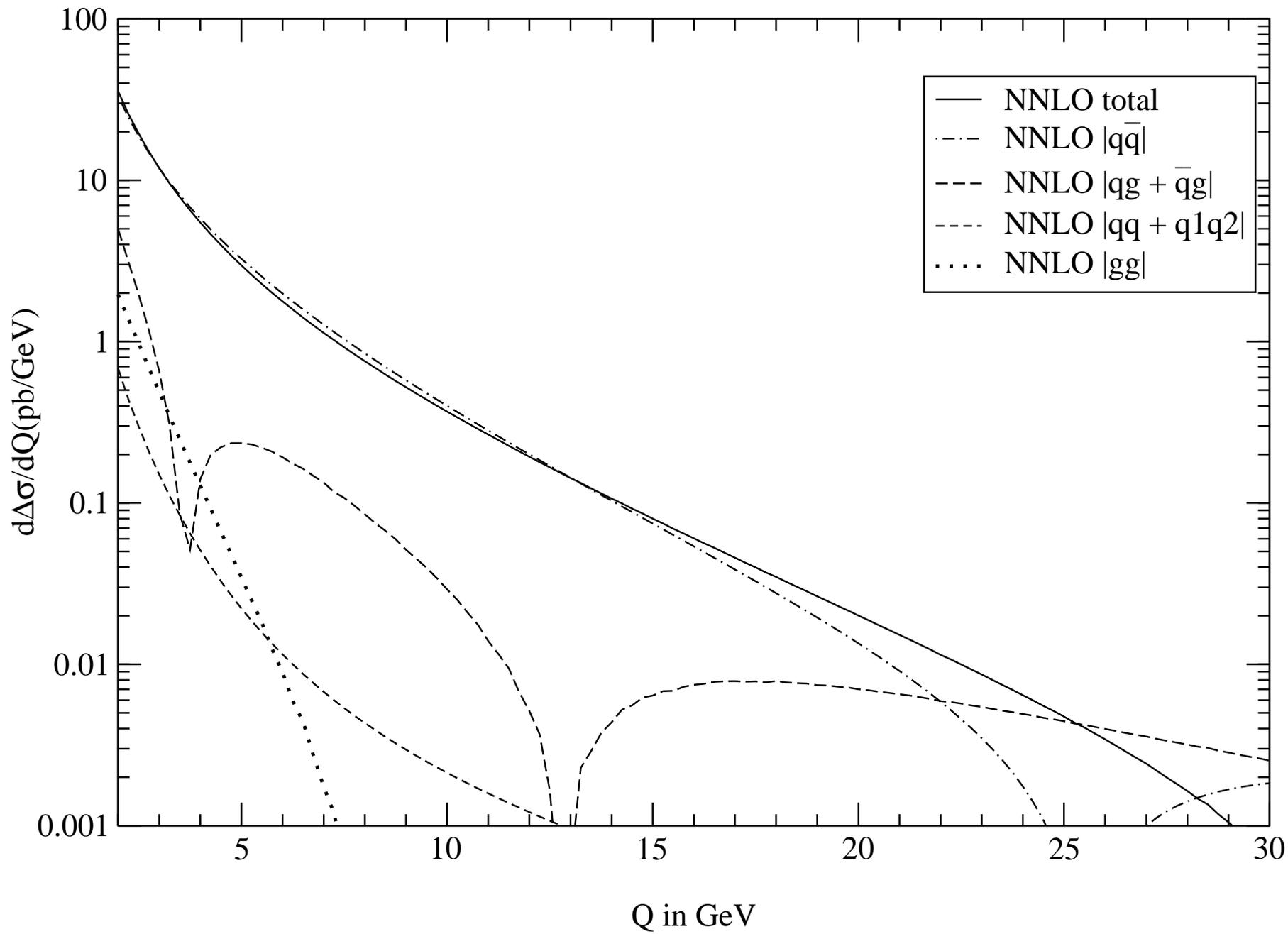
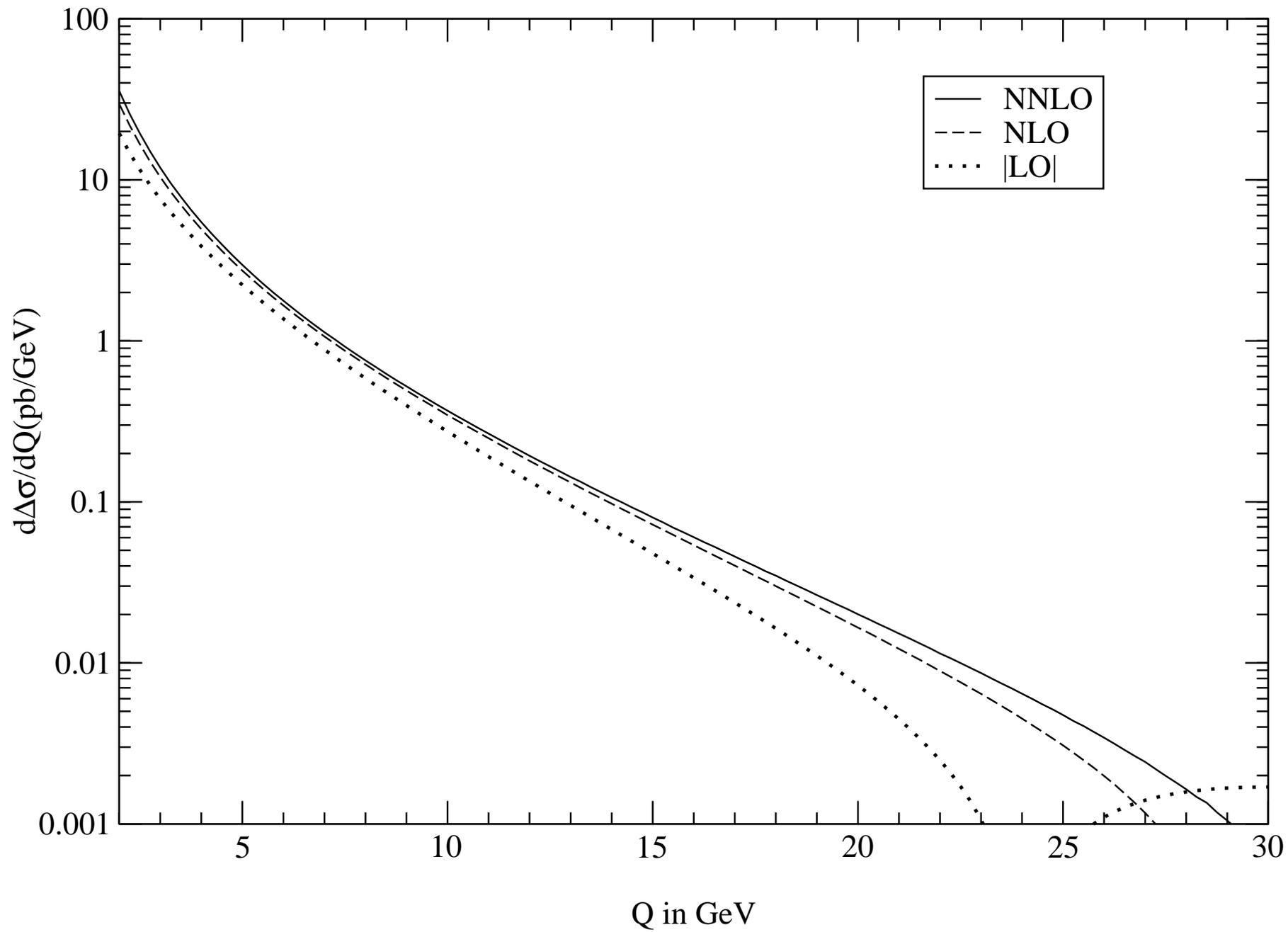
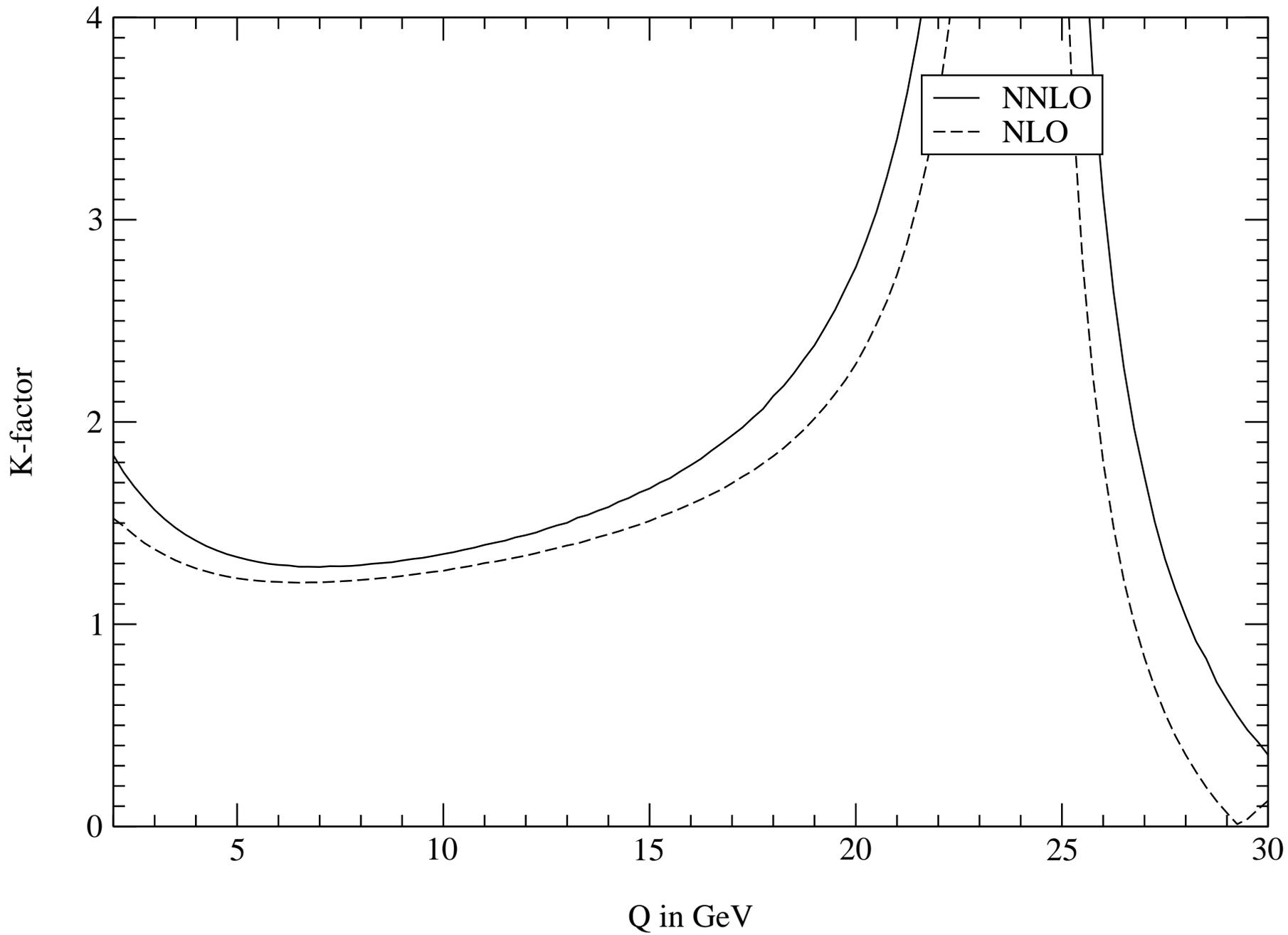
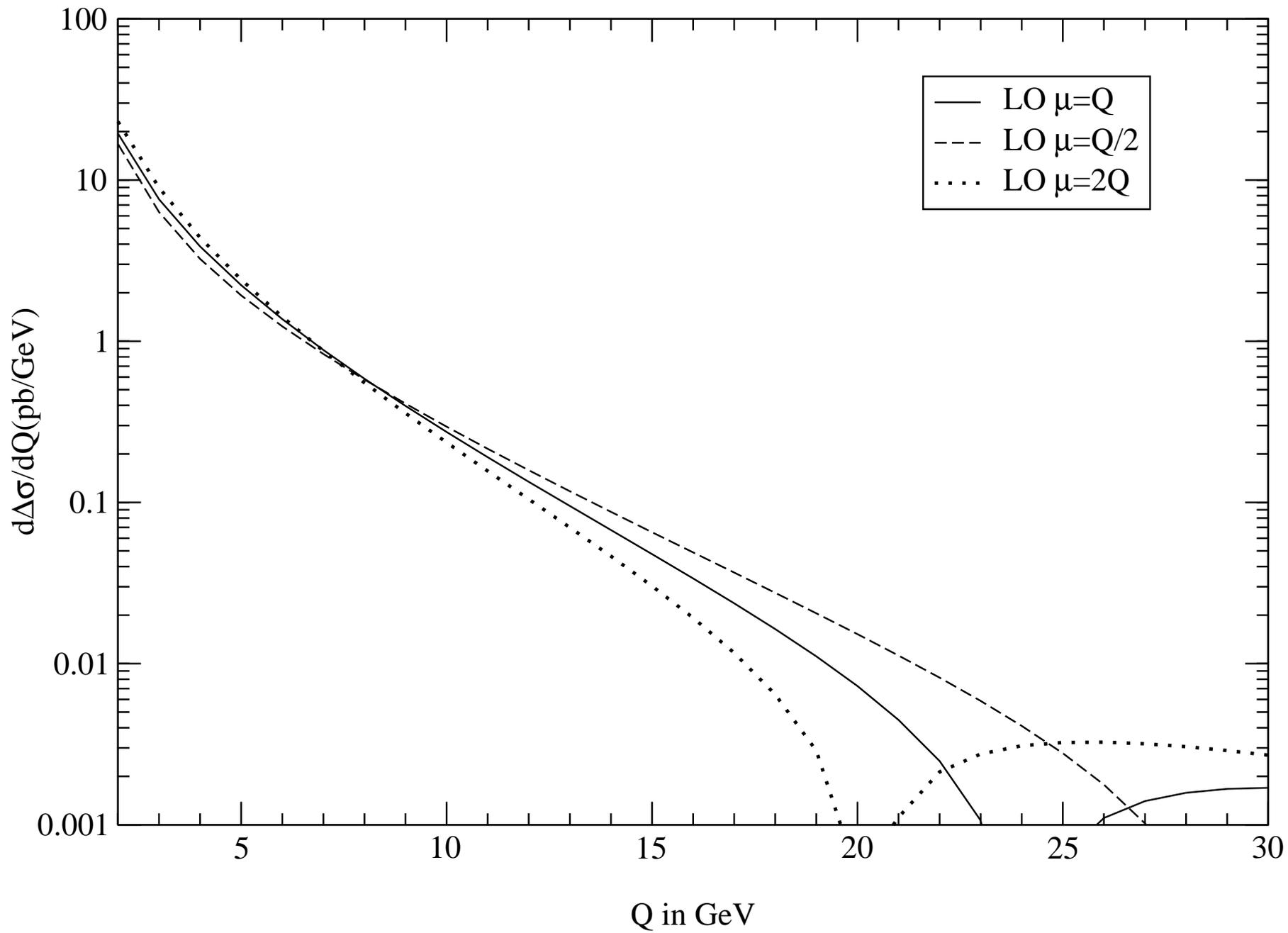
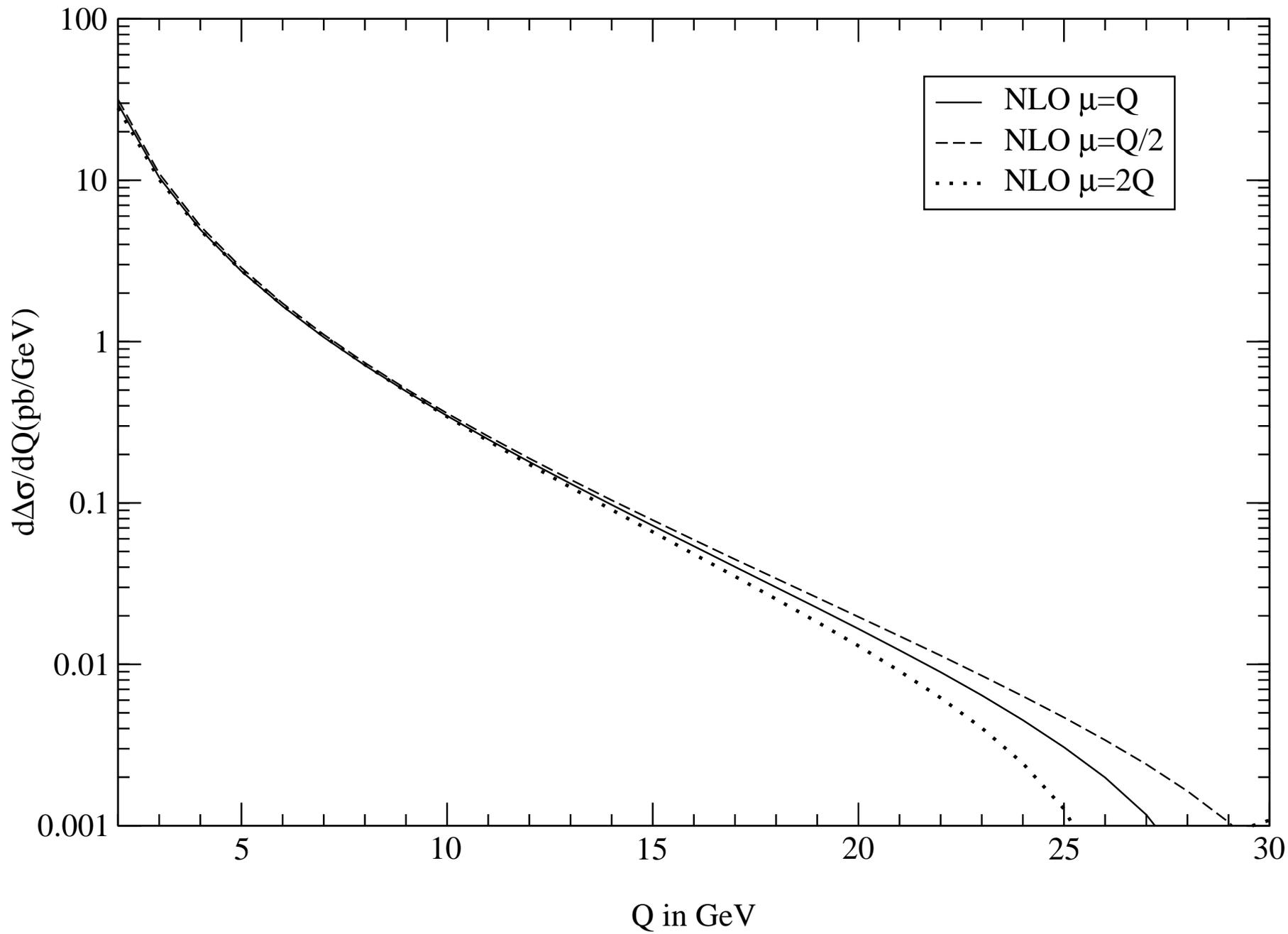
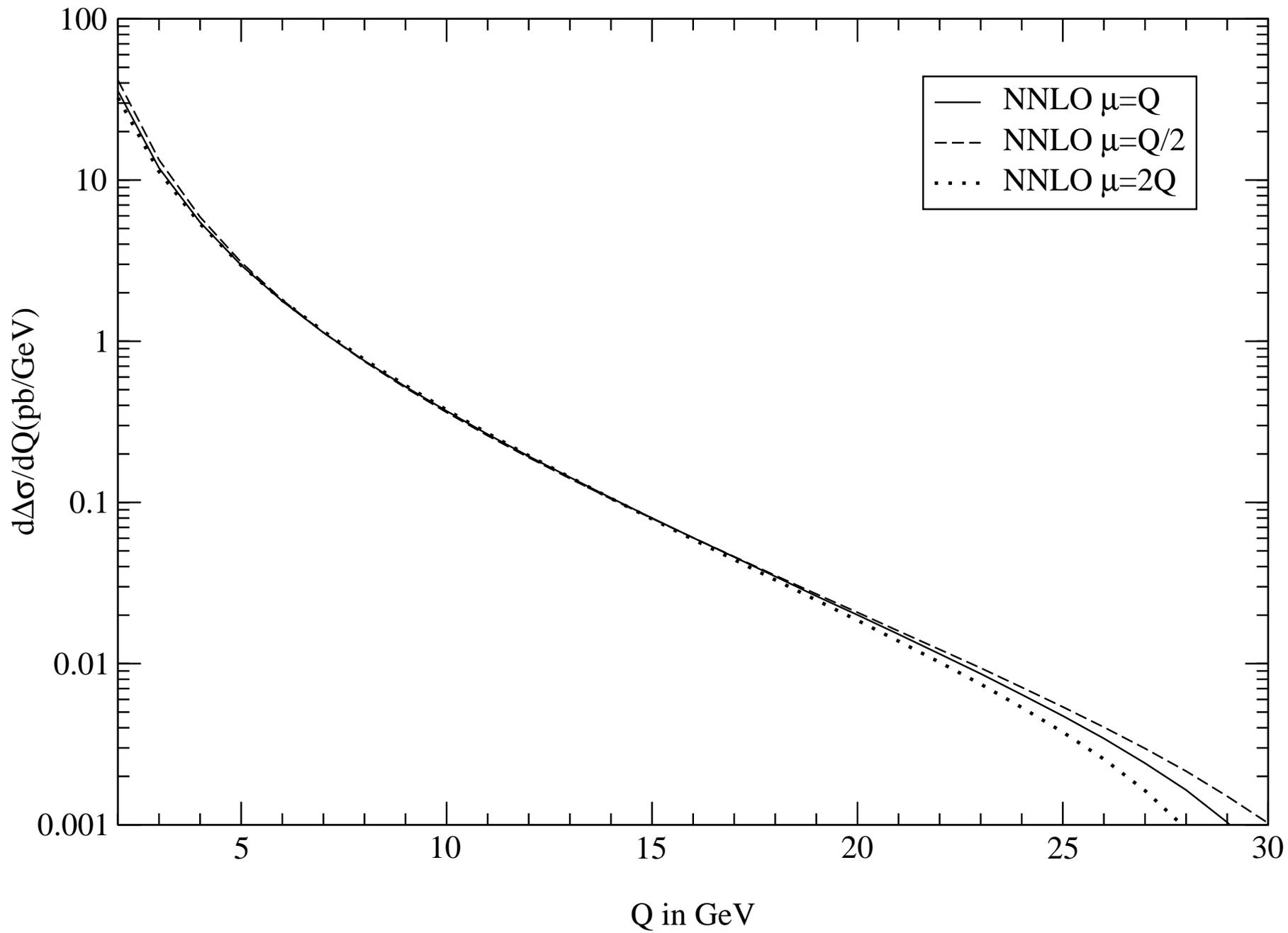
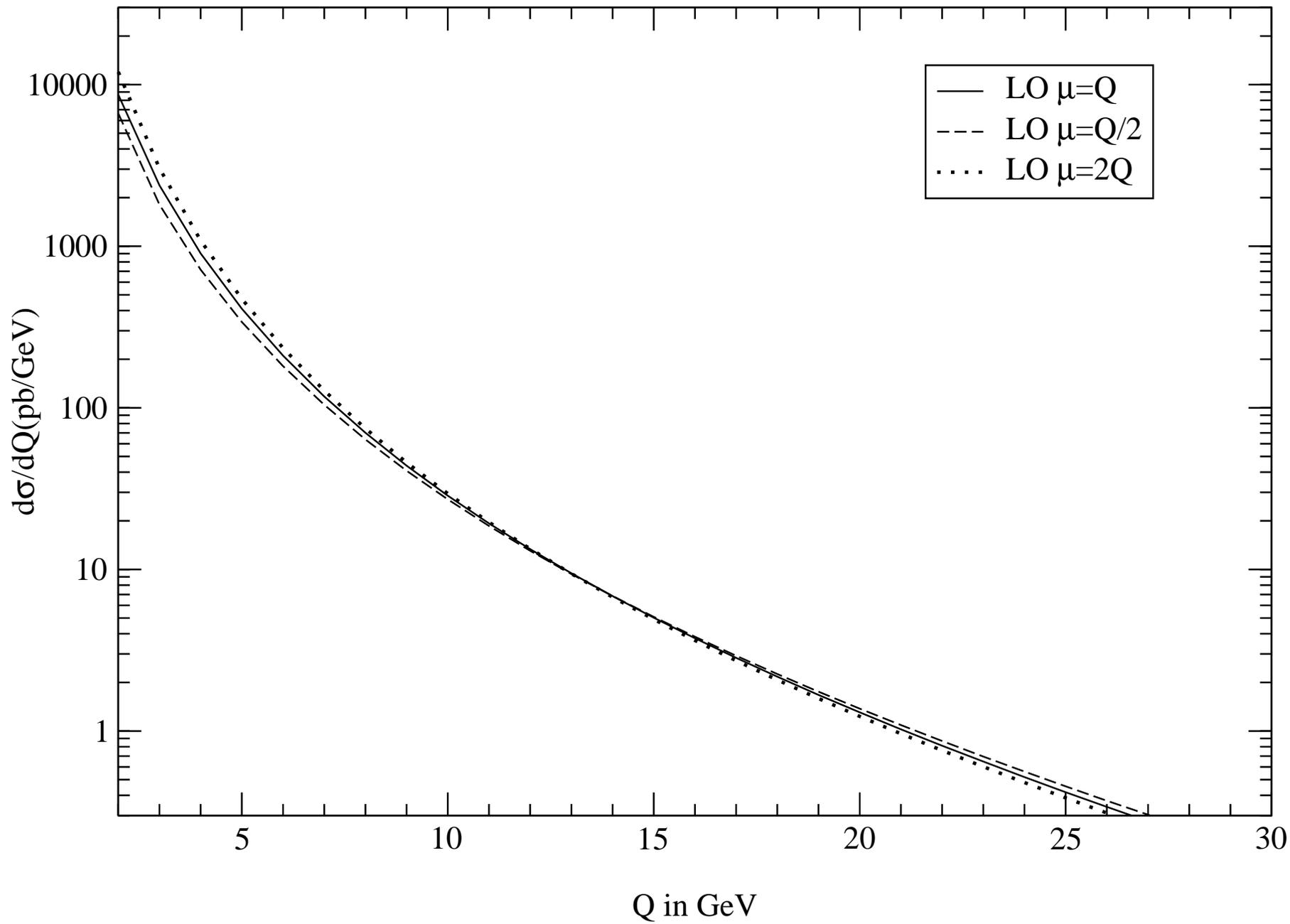
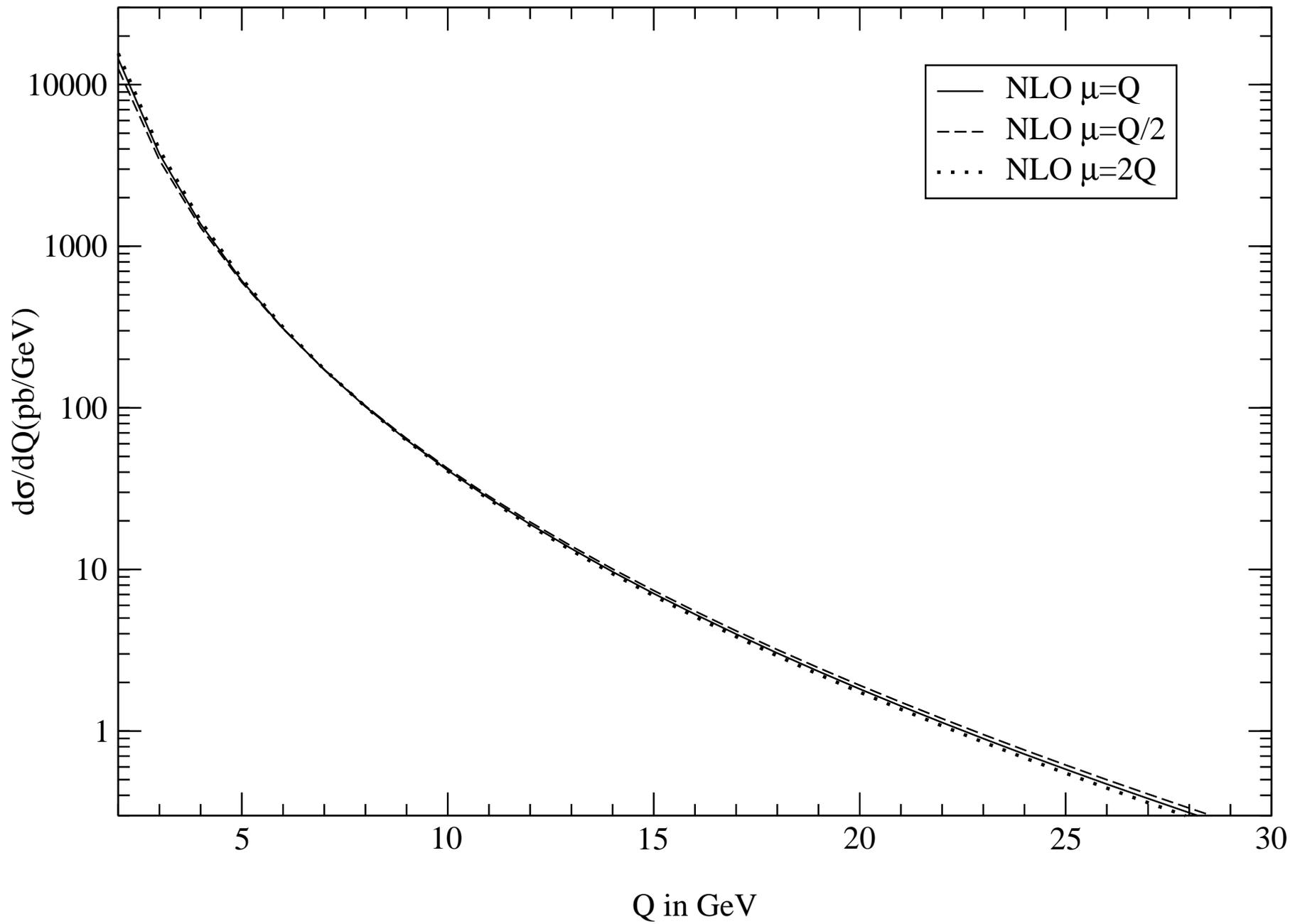
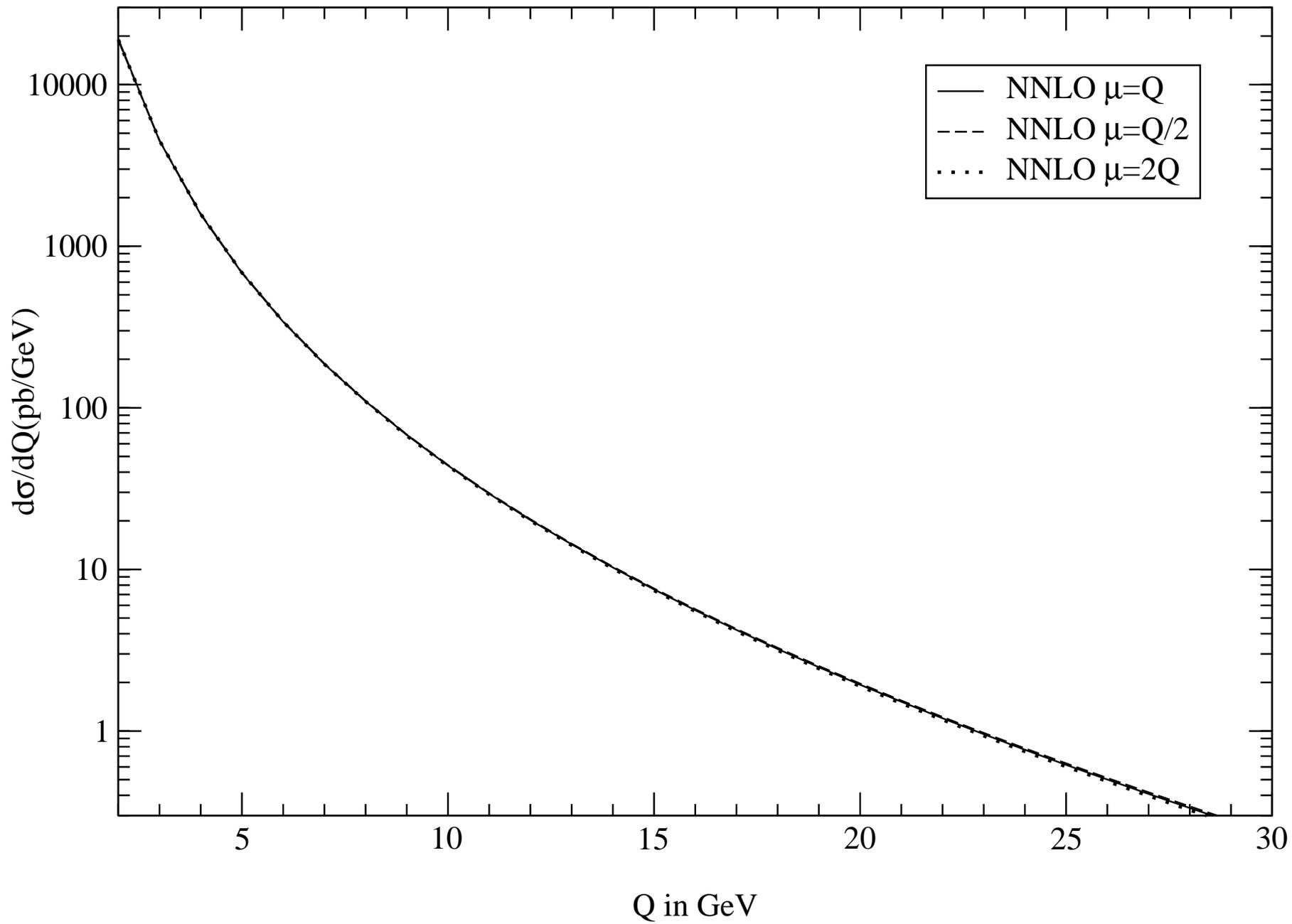
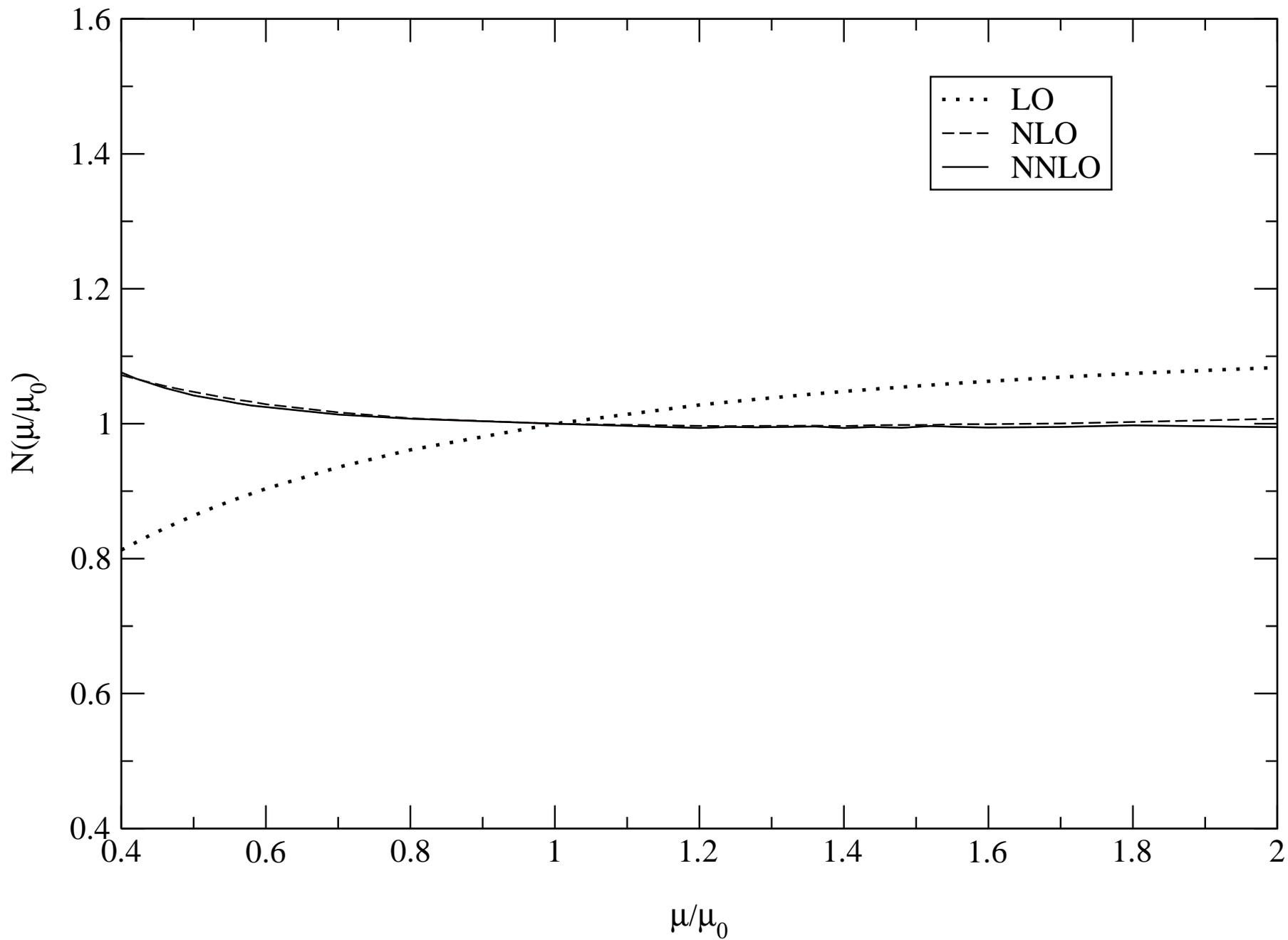
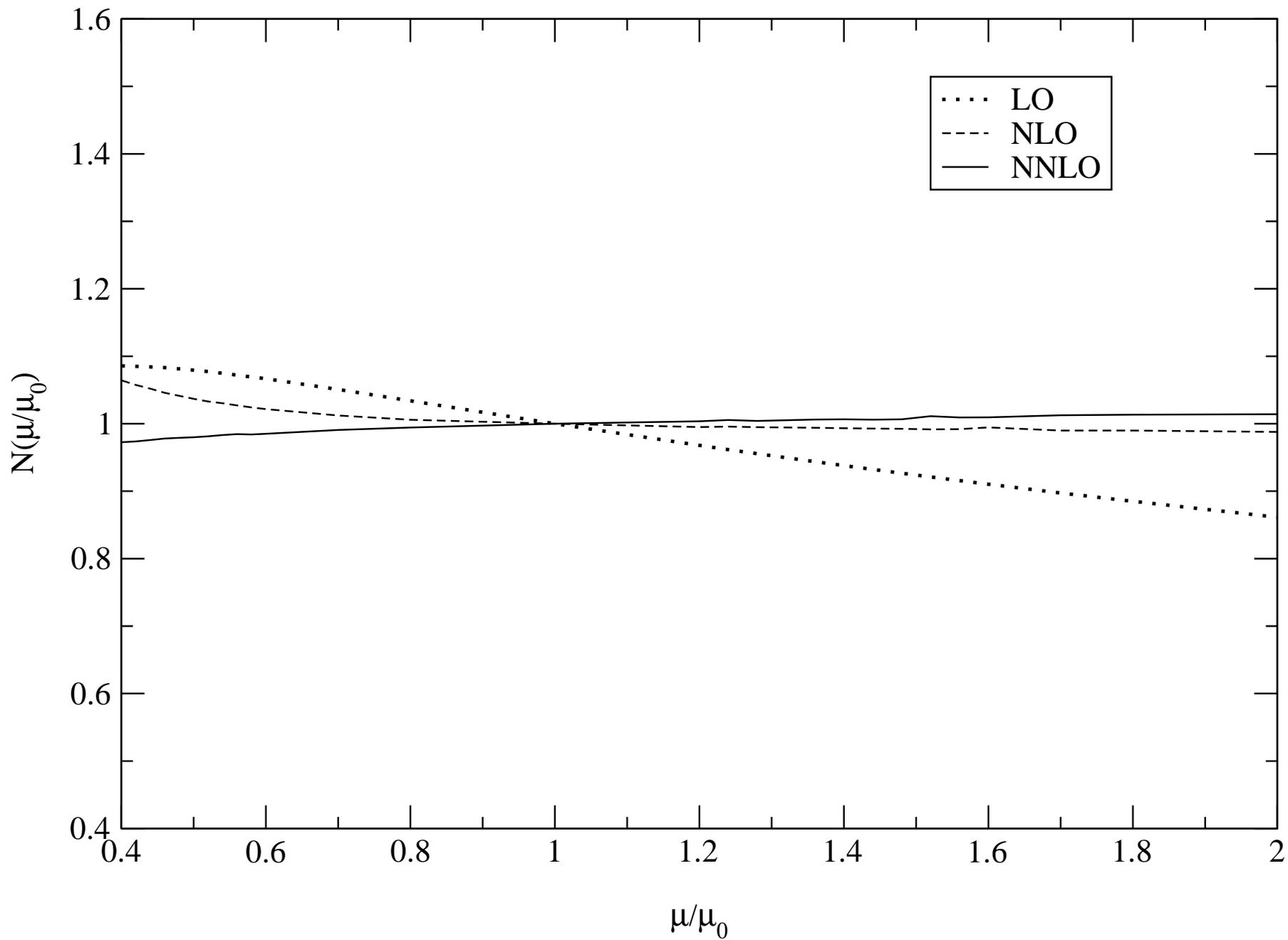
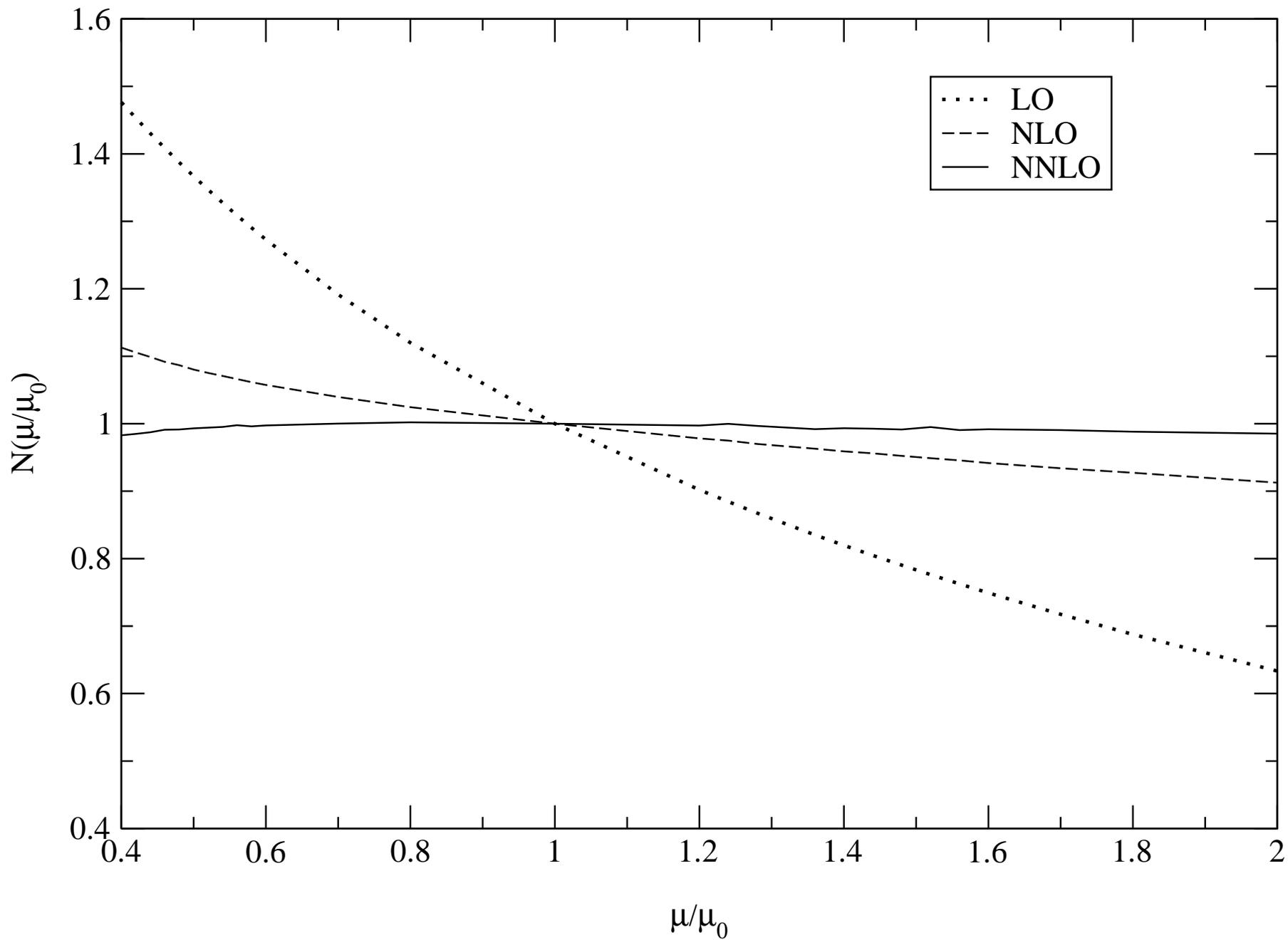
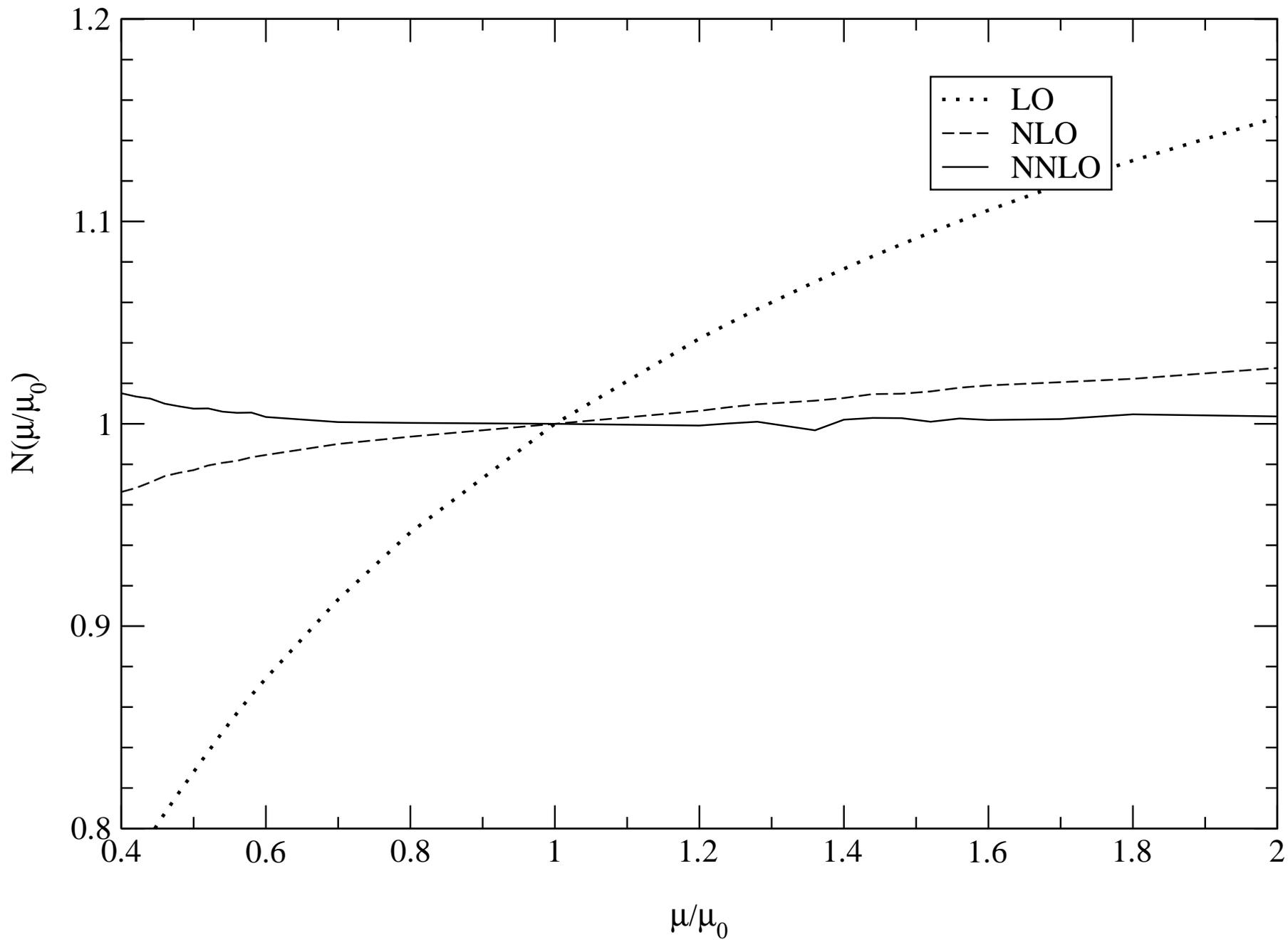
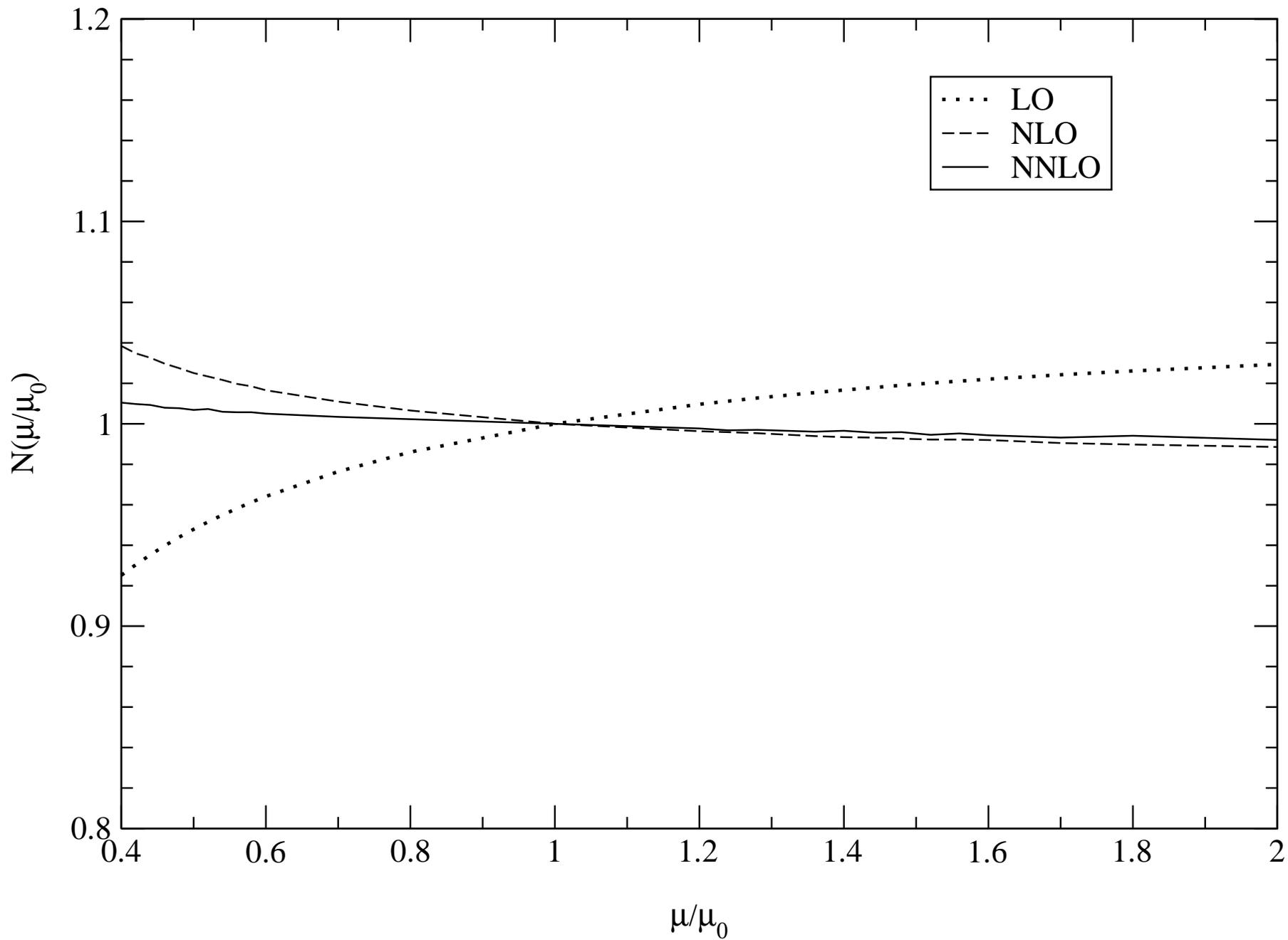
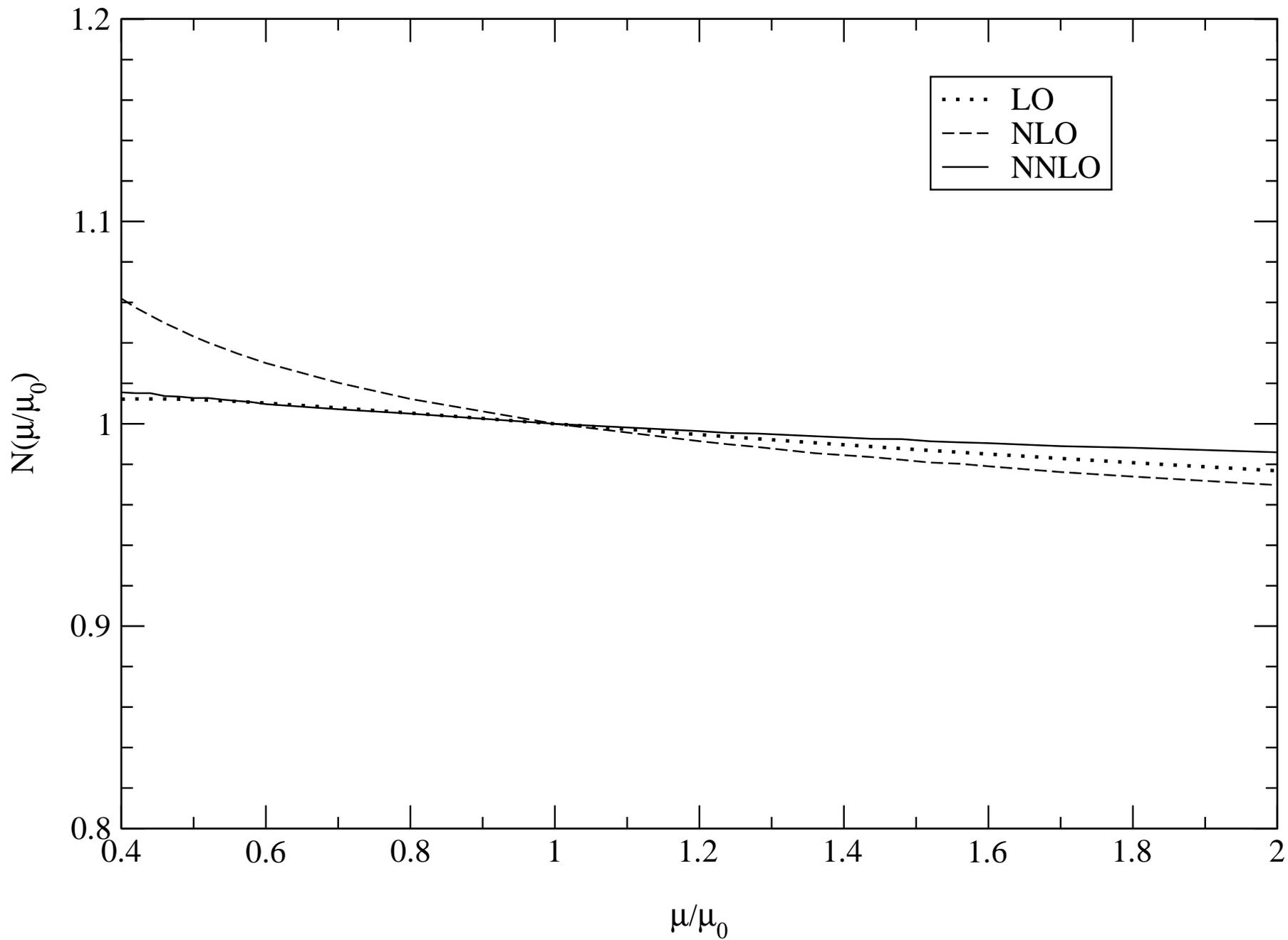
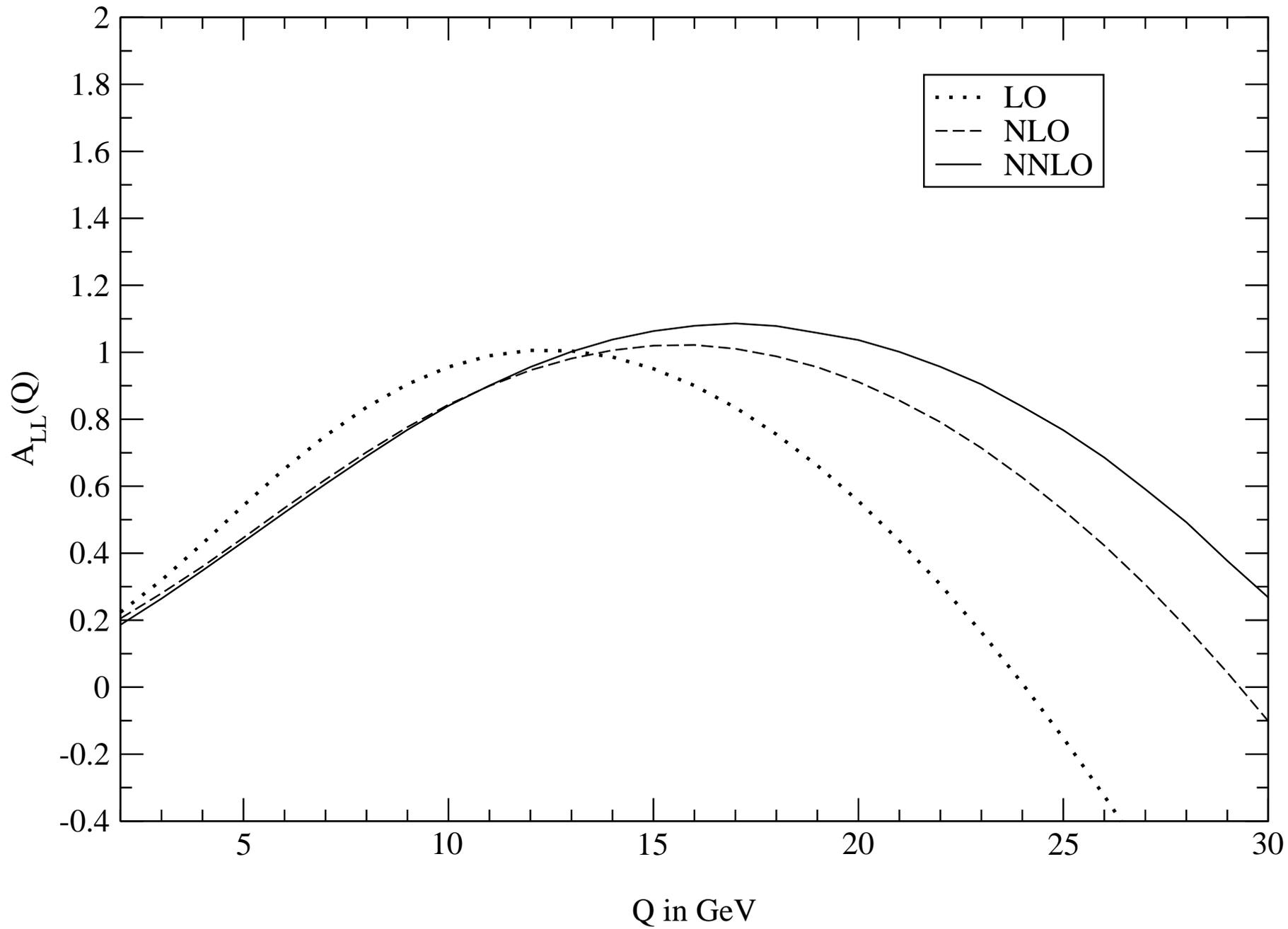
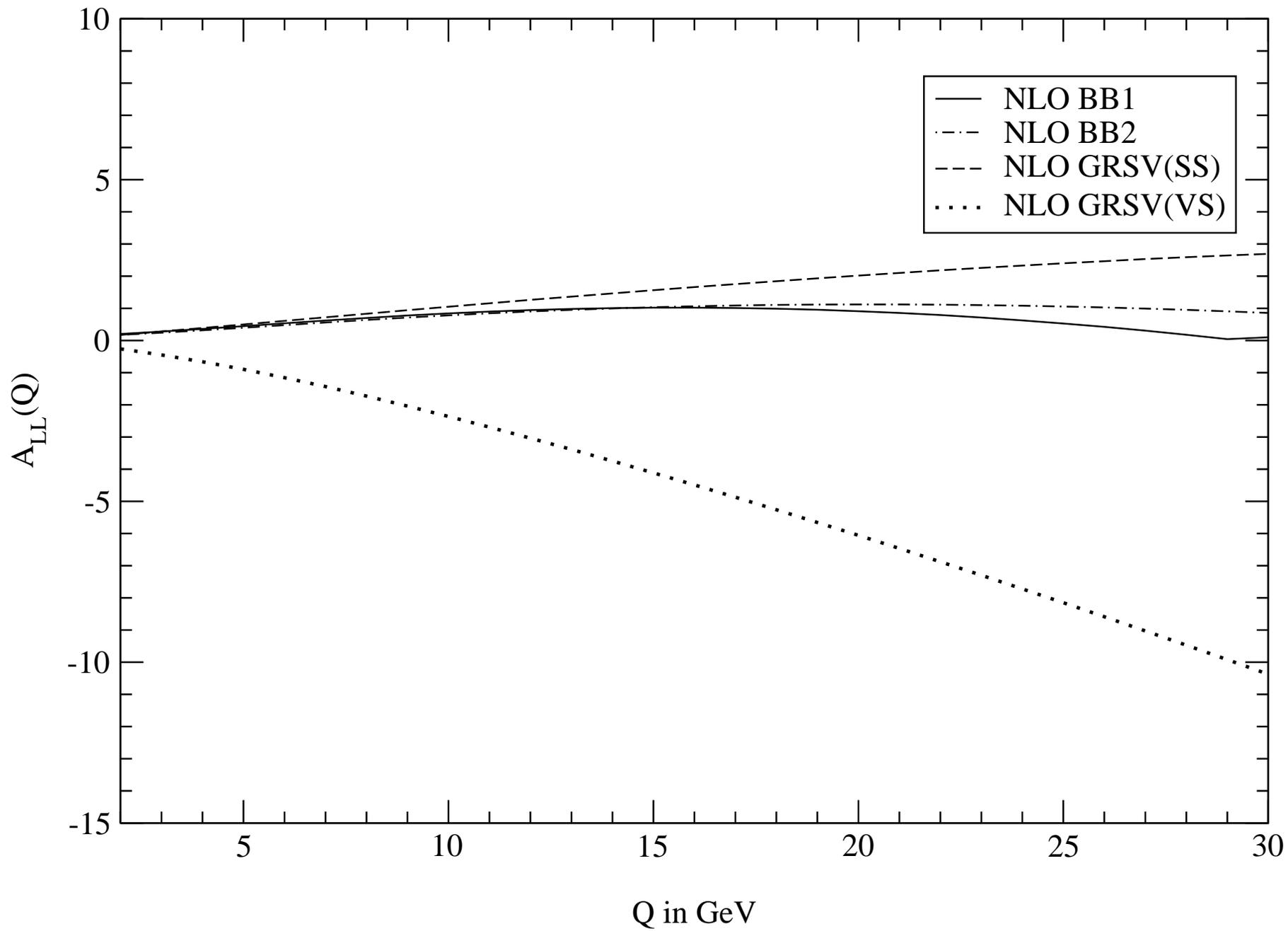
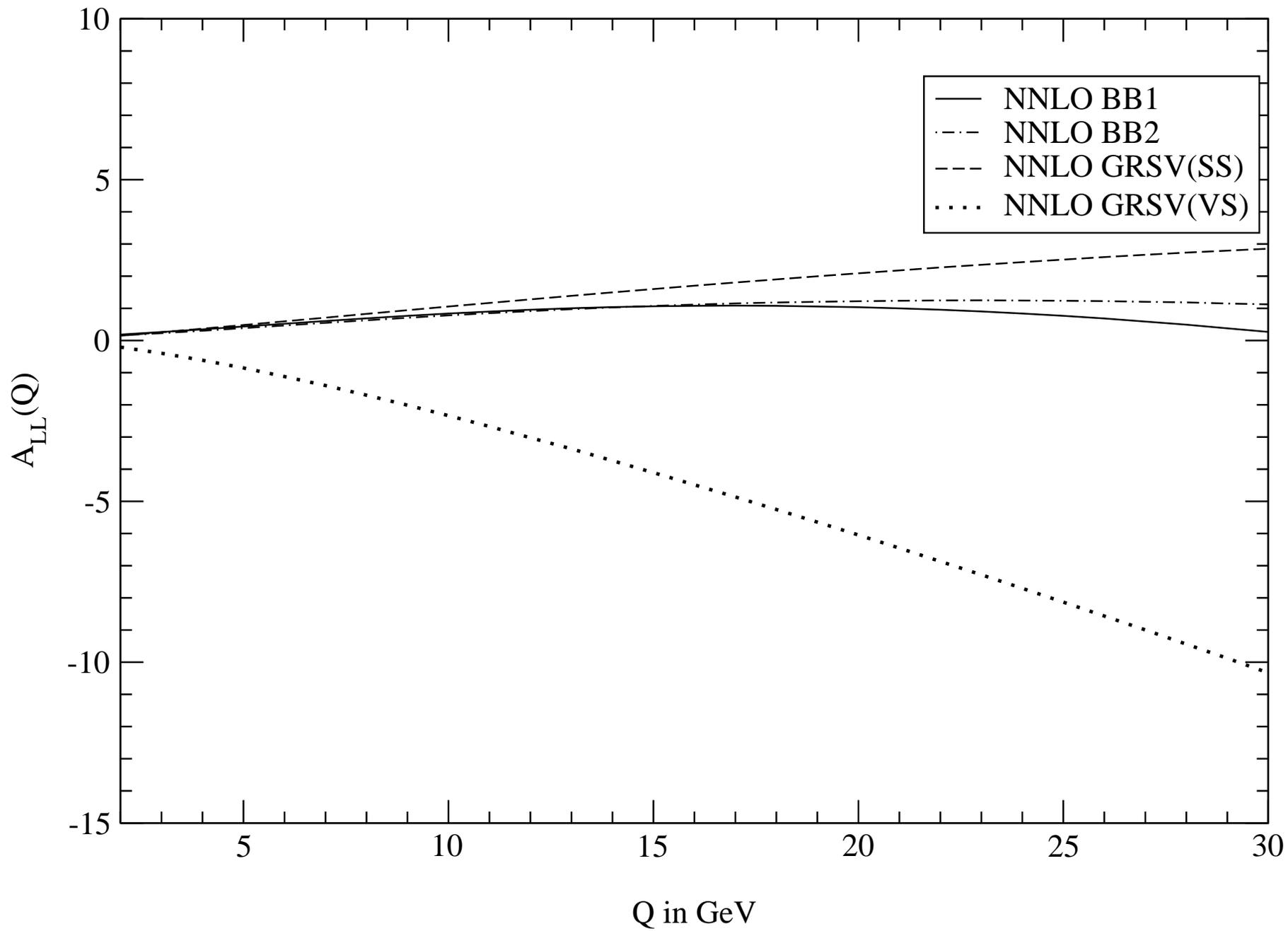
\end{document}